\documentclass[twocolumn,           
               showpacs,            
               preprintnumbers,     
               aps,                 
               prd,                 
               a4paper,             
               superscriptaddress,  
               nofootinbib,         
               tightenlines,        
               floats,floatfix      
               ]{revtex4}

\usepackage{ graphicx }
\usepackage{ latexsym }
\usepackage{ amsmath, amssymb }        
\usepackage[ draft=false ]{ hyperref }
\usepackage{multirow}

\newcommand{\rd}{\mathrm{d}}
\newcommand{\DLie}{\mathcal{L}}
\newcommand{\R}{\mathbb{R}}
\newcommand{\e}{\mathrm{e}}

\begin{document}


\title{Numerical evolution of multiple black holes with accurate initial data}
\author{Pablo  Galaviz}   \author{Bernd  Br\"ugmann}  \affiliation{Theoretical
  Physics Institute, University of Jena, 07743 Jena, Germany} \author{Zhoujian
  Cao} \affiliation{Institute  of Applied Mathematics,  Academy of Mathematics
  and Systems Science, Chinese Academy of Sciences, Beijing 100190, China}


\date{June 2, 2010}


\begin{abstract}
  We present  numerical evolutions of  three equal-mass black holes  using the
  moving  puncture approach.   We calculate  puncture initial  data  for three
  black  holes solving  the  constraint  equations by  means  of a  high-order
  multigrid elliptic  solver.  Using these  initial data, we show  the results
  for three  black hole evolutions with sixth-order  waveform convergence.  We
  compare results obtained with  the \textsc{BAM} and \textsc{AMSS-NCKU} codes
  with previous results.  The approximate analytic solution to the Hamiltonian
  constraint  used in  previous  simulations  of three  black  holes leads  to
  different  dynamics and  waveforms.  We  present some  numerical experiments
  showing the  evolution of four  black holes and the  resulting gravitational
  waveform.
\end{abstract}




\maketitle



\section{Introduction}
\label{sec:introduction}

The gravitational $n$-body problem is an old and important problem which dates
back to  1687, when Isaac  Newton's ``Principia'' was published.   The special
case for  $n=3$ was studied  by Euler, Lagrange, Laplace,  Poincar\'{e}, among
others  (see, e.g.,~\cite{ValHan06}).   However, solutions  of  the three-body
problem  have  shown a  rich  complexity and  are  far  from being  completely
understood.  From  the point  of view of  celestial mechanics,  the three-body
problem is  related to the  important question of  the stability of  the solar
system \cite{Las89,Las94,Hay07}.   In globular clusters  $n$-body interactions
appear  to be  important in  the  formation of  intermediate-mass black  holes
\cite{Gul03,ColMil03,GulMilHam04,ValMik91,ZwaMcM00,GuaZwaSip05,PreBerBer08}.
Finally,  covering  large  scales  in  cosmology numerical  solutions  of  the
$n$-body   problem    are   used   to   simulate    formation   of   structure
\cite{VolEtal05,Ber98,Hat03,FreCheZwa04}.

In  order   to  solve  the  three-body  problem,   scientists  have  developed
mathematical tools,  and with the  development of computers since  the 1950's,
also numerical  techniques.  On the  one hand, there are  analytical solutions
for special  cases of the three-body  problem, for example those  due to Euler
and Lagrange (see e.g.~\cite{ValHan06,GolPooSaf01}),  on the other hand, there
are solutions which exhibit a chaotic  behavior.  At the beginning of the 20th
century  Sundman  found a  convergent series  solution to  the three-body
problem \cite{Sun1907,Bar96}.  However, the  rate of convergence of the series
which he  had derived is  extremely slow, and  it is not useful  for practical
purposes.  From the point of view of dynamical systems, the three-body problem
was a  key system  which allowed  Poincar\'{e} to identify  many of  the novel
ideas related to the dynamical system theory and chaos \cite{Bar96}.
 
Using  post-Newtonian techniques  (PN), it  is  now possible  to describe  the
dynamics   of   $n$    compact   objects,   up   to   3.5    PN   order   (see
e.g.~\cite{JarSch97,Bla02,FutIto07}).  For binary  systems the ADM Hamiltonian
has  been specialized up  to 3.5  PN order  \cite{KonFaySch03}, and  for three
bodies    there    are    explicit    formulas    up    to    2    PN    order
\cite{Chu09,Sch87,LouHir08}.  Periodic solutions,  also known as choreographic
solutions,        were       studied       using        these       techniques
\cite{Moo93,TatTakHid07,LouHir08}, as  well as estimates  of the gravitational
radiation              for              binary-single             interactions
\cite{GulMilHam05,ZwaMcM00,CamDetHan05}.

The first complete simulations using general-relativistic numerical evolutions
of  three black  holes  were presented  in \cite{CamLouZlo07f,LouZlo07a}  (see
\cite{Bru97a,Die03,Bru97atext}  for very  limited early  examples  of multiple
black  hole simulations).   \nocite{Bru96,Bru97,emc2}  The recent  simulations
show  that  the  dynamics  of  three compact  objects  display  a  qualitative
different behavior than the Newtonian dynamics.
Gravitational  waves are  an  extra  component in  the  three-body problem  of
compact objects which enrich the  phenomenology of the system.  The changes in
the energy and  momentum resulting from the gravitational  radiation produce a
difference in the dynamics of the  system. There are open questions related to
the  general-relativistic dynamics  of $n$  compact objects,  for  example the
possible  chaotic behavior of  the dynamics  of $n$  black holes,  the inverse
problem  in gravitational  wave  emission, the  existence of  quasi-stationary
solutions and their stability, etc.

In some regard simulations of three  or more black holes are more sensitive to
small  changes  in  the  data  than  binary  simulations,  as  also  noted  in
e.g.\ \cite{CamLouZlo07f,LouZlo07a}. In a typical binary, changing the initial
momentum slightly leads to a  correspondingly small change in the eccentricity
of  the binary.  In a  black hole  triple, changing  the initial  momentum may
change the dynamics  completely since the first black hole  can merge with the
second or the  third depending on the initial momentum.  Hence, moving up from
two to three bodies introduces a new  feature, but of course from the point of
view of $n$-body simulations and chaotic systems this is no surprise.

In this paper we examine  the sensitivity of the fully relativistic evolutions
of three black holes to changes in the initial data. We present simulations of
three and four black holes. The examples for three black holes are some of the
simpler  cases  already considered  in  \cite{CamLouZlo07f,LouZlo07a}. A  more
detailed analysis about a possible  chaotic behavior of the three body problem
in general relativity is beyond the scope of this work, but would certainly be
of interest.

In \cite{CamLouZlo07f,LouZlo07a}, initial data is specificed using an analytic
approximation, which introduces a finite  error that does not converge to zero
with  numerical resolution.  The  reason to  use  such initial  data is  that,
although accurate initial data for  two black holes is readily available, this
is not the case for more than  two black holes. Below we show that solving the
constraints  numerically to  obtain initial  data for  an arbitrary  number of
black holes, the result of  the evolutions can change dramatically. The actual
difference between  the analytic approximation and the  numerical initial data
is not  large (depending  on the initial  parameters), but, as  expected, even
small differences can lead to large changes for multiple black hole orbits.

The paper is organized as follows. In Sec.~\ref{sec:initialdata} we review the
puncture method \cite{BeiOMu94,BeiOMu96,BraBru97}, which is the basic approach
that  we use  to  solve  the initial  data  problem.  This  is  followed by  a
description  of  our  new  code,  \textsc{Olliptic},  designed  to  solve  the
constraint  equations  of the  3+1  formalism numerically.   \textsc{Olliptic}
implements a parallel multigrid algorithm  on nested regular grids, with up to
eighth order  finite differencing.  In Sec.~\ref{sec:results},  we present our
results for three  test cases, for the initial data of  a single puncture, and
for two and three punctures.  The  evolution of three black holes is presented
in  Sec.~\ref{sec:num_evol_3bhs},   where  we  compare   \textsc{BAM}  results
directly with the \textsc{AMSS-NCKU} code \cite{CaoYoYu08} and indirectly with
the  previous results of  \cite{CamLouZlo07f,LouZlo07a}.  Finally,  we perform
simulations of four black holes to show that the same techniques work for more
than  three  back  holes  (our  largest simulation  involves  42  black  holes
\cite{emc2}). We conclude with a discussion in Sec.~\ref{sec:discussion}.


\section{INITIAL DATA}
\label{sec:initialdata}

Under  a  3+1  decomposition, the  Einstein  equations  split  into a  set  of
evolution  equations  and constraint  equations,  namely  the Hamiltonian  and
momentum constraints  (see, e.g.,~\cite{Alc08,Coo00,Gou07b} for  reviews).  In
vacuum the constraint equations read as follows:
\begin{eqnarray}
  \nabla_j \left( K^{ij} - \gamma^{ij} K \right) &=& 0,\label{eqn:MomConst} \\  
  R^2 + K^2 - K_{ij} K^{ij} &=& 0, \label{eqn:HamConst}
\end{eqnarray}
where $R$ is the Ricci scalar, $K_{ij}$ is the extrinsic curvature and $K$ its
trace, $\gamma_{ij}$  is the 3-metric, and $\nabla_j$  the covariant derivative
associated with $\gamma_{ij}$.


\subsection{Puncture method}
\label{sec:punctures}

The  constraints can  be  solved, for  example,  with the  puncture method  of
\cite{BraBru97}.  $N$ black holes  are modeled by adopting the Brill-Lindquist
wormhole topology  \cite{BriLin63} with  $N$+1 asymptotically flat  ends which
are compactified and  identified with points $r_i$ on  $\R^3$.  The coordinate
singularities at the points $r_i$ resulting from compactification are referred
to as punctures.

Following the  conformal transverse-traceless decomposition  approach, we make
the following assumptions for the metric and the extrinsic curvature:
\begin{eqnarray}
  &\gamma_{ij} =  \psi_0^4 \tilde{\gamma}_{ij},& \label{eqn:ConfGamma} \\   
  &K_{ij} = \psi_0^{-2} \tilde{A}_{ij} 
  + \frac{1}{3} K \gamma_{ij},&\label{eqn:ConfK}
\end{eqnarray}
where $\tilde{A}^{ij}$ is trace free.   We choose an initially flat background
metric,  $\tilde{\gamma}_{ij}=\delta_{ij}$, and a  maximal slice,  $K=0$.  The
last            choice           decouples            the           constraint
equations~(\ref{eqn:MomConst})-(\ref{eqn:HamConst}) which take the form
\begin{eqnarray}
  &  \partial_j \tilde{A}^{ij} = 0, & \label{eqn:MomCTT} \\ 
  & \vartriangle \psi_0 + \frac{1}{8} \tilde{A}^{ij} \tilde{A}_{ij} 
  \psi_0^{-7}=0. & \label{eqn:HamCTT}
\end{eqnarray}

Bowen  and  York  \cite{BowYor80}  have  obtained a  non-trivial  solution  of
Eq.~(\ref{eqn:MomCTT})  in a  Cartesian  coordinate system  ($x^i$), which  by
linearity of the momentum constraint can be superposed for any number of black
holes (here the index $n$ is a label for each puncture):
\begin{eqnarray}
  \tilde{A}^{ij} &=& \sum_n \left[ \frac{3}{2 r^3_n} \left[ x_n^i P_{n}^{j}   
      + x_n^j P_{n}^{i} - \left(  \delta^{ij} 
      - \frac{ x_n^i x_n^j}{r_n^2} \right) P^n_k x_n^k \right]
    \right.\nonumber \\&& \left.
    + \frac{3}{r_n^5} \left(  \epsilon^{ik}_{\;\;l} S^n_k x_n^l  x_n^j +
    \epsilon^{jk}_{\;\;l} S^n_k x_n^l x_n^i \right) \right],
\end{eqnarray}
where   $r_n:=\sqrt{x_n^2+y_n^2+z_n^2}$,    $\epsilon^{ik}_{\;\;l}$   is    the
Levi-Civita tensor  associated with the flat  metric, and $P_i$  and $S_i$ are
the ADM linear and angular momentum, respectively.

The Hamiltonian constraint (\ref{eqn:HamCTT}) becomes an elliptic equation for
the conformal factor. The solution is split  as a sum of a singular term and a
finite correction $u$ \cite{BraBru97},
\begin{equation}
  \psi_0 = 1 + \sum_n \frac{m_n}{2r_n} + u, \label{eqn:ConfPunct}
\end{equation}
with  $u \rightarrow  0$ as  $r_n \rightarrow  \infty$.  The  function  $u$ is
determined  by an  elliptic equation  on $\R^3$  and is  $C^\infty$ everywhere
except at the punctures, where it  is $C^2$. The parameter $m_n$ is called the
bare mass of the $n$th puncture.


\subsection{Numerical Method}
\label{sec:numerical_method}

In  order  to  solve   Eq.~(\ref{eqn:HamCTT})  numerically,  we  have  written
\textsc{Olliptic}, a  parallel computational  code to solve  three dimensional
systems of  non-linear elliptic equations with  a 2nd, 4th, 6th, and 8th order
finite difference multigrid method.   The elliptic solver uses vertex-centered
stencils  and box-based mesh  refinement that we  describe  below.   We use  a
standard  multigrid  method  \cite{Bra77,BaiBra87,BraLan88,HawMat03,ChoUnr86b}
with a Gauss-Seidel Newton relaxation algorithm (e.g.~\cite{Cho06a}).

The  numerical  domain is  represented  by  a  hierarchy of  nested  Cartesian
grids. The hierarchy consists of $L+G$ levels of refinement indexed by $l = 0,
\ldots ,L + G - 1$. A refinement level consists of one or more Cartesian grids
with constant grid-spacing $h_l$ on level  $l$.  A refinement factor of two is
used such that  $h_l = h_G/2^{|l-G|}$.  The grids are  properly nested in that
the coordinate extent  of any grid at  level $l > G$ is  completely covered by
the grids at  level $l-1$.  The level $l=G$ is the  ``external box'' where the
physical  boundary is  defined.   We use  grids  with $l<G$  to implement  the
multigrid method beyond level $l=G$.

\begin{figure}[ptb]
  \centering \includegraphics[scale=0.25]{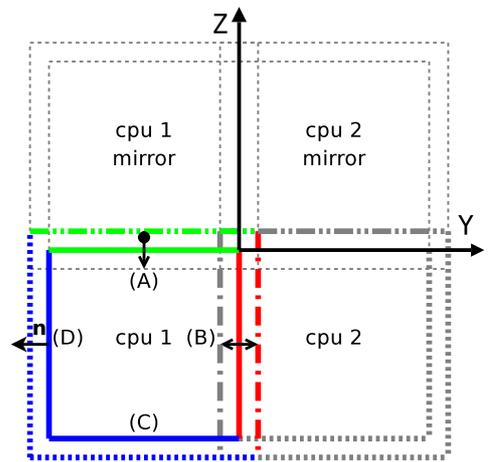}
  \caption{2D representation of a domain divided between 2 processors and with
    bitant symmetry.  Face (A) is internal  to the global domain and handles a
    reflection symmetry.  Face  (B) is internal to the  domain and manages the
    communication with the  second processor.  Faces (C) and  (D) are physical
    boundaries,  where we  impose a  Robin  boundary condition  in the  normal
    direction $\vec{n}$ (see text).}
  \label{fig:GridDia}
\end{figure}

The parallelization approach that we use is block decomposition, in which each
domain is divided into rectangular  regions among the processors such that the
computational work  load is balanced. For  levels $l \ge G$  every domain uses
$p/2$ buffer  points at  the boundary  of the domain  (here $p$  indicates the
order of the  finite difference stencil).  Levels with  $l<G$ contain a single
point at  the boundary.  For every  face of the  three dimensional rectangular
domain    we    use    these    points    for    different    purposes    (see
Fig.~\ref{fig:GridDia}):
\begin{enumerate}
\item If the face is on the outside of the global domain, we use the points as
  a  refinement  boundary  (or  physical  boundary  if  $l=G$);  the  boundary
  conditions are explained below.
\item If the  face is in the internal  part of the global domain,  then we use
  ghost  zones of  the neighboring  processors  to update  information of  the
  buffer points.
\item If the  face is defined with symmetry, we use  a reflection condition to
  calculate the values at the boundary.
\end{enumerate}
\textsc{Olliptic}    can    be   used    with    three   symmetries:    octant
$(-x,-y,-z)\rightarrow(x,y,z)$,   quadrant  $(-x,-y,z)\rightarrow(x,y,z)$  and
bitant $(x,y,-z)\rightarrow(x,y,z)$. We use the negative part of the domain to
define the computational  grid, because that increases the  performance of the
relaxation  method somewhat  since the  resulting order  of  point traversal
helps propagating boundary information into the grid.

For the ``physical'' or outer boundary we require that $u \rightarrow A$ as $r
\rightarrow \infty$. The standard condition used  in this case  is an inverse
power fall-off,
\begin{equation}
  u(r)=A+\frac{B}{r^q}, \quad \mathrm{for} \quad r \gg 1, \; q > 0, 
  \label{eq:DirichBC}
\end{equation} 
where the factor $B$ is unknown. It is possible to get an equivalent condition
which   does   not   contain    $B$   by   calculating   the   derivative   of
(\ref{eq:DirichBC})  with respect  to $r$,  solving the  equation for  $B$ and
making a substitution in the original equation. The result is a \textit{Robin}
boundary condition:
\begin{equation}\label{eq:RobinBC}
  u(\vec{x}) +  \frac{r}{q} \frac{\partial u(\vec{x})}{\partial r} = A.
\end{equation} 

The implementation of the boundary condition was a key point
to get accurate solutions, so we describe our implementation
in some detail.
Rather than taking derivatives in the radial direction as is required
by (\ref{eq:RobinBC}), we take derivatives only in the direction 
normal to the faces of our rectangular domain. At the edges of the
boundary, we use a linear combination of the derivatives
along the normals of the two adjacent faces. At the corners, we use 
a linear combination of the derivatives for the three adjacent faces. 
In the computation, we first apply the boundary condition to the
interior of the boundary faces, then compute derivatives inside the
faces to update the edges, and then compute derivatives inside the
edges to obtain boundary data at the corners.
We use a one sided finite difference stencil
of  order $p$  and  a Newton  iteration method  to  update the  values on  the
boundary.  For example,  for the  face (D)  (see Fig.~\ref{fig:GridDia}), the
equations are
\begin{eqnarray}
  R_{ijk} &:=&  D_y^{p+} u^n_{ijk}  
  -\frac{q  y_{ijk}}{r_{ijk}^2} (A-u^n_{ijk}), \label{eqn:ResBoundFD}\\ 
  \rd R_{ijk} &:=& \frac{\partial}{\partial u_{ijk}} D_y^+ u^n_{ijk}  
  + \frac{q y_{ijk}}{r_{ijk}^2} u^n_{ijk},\\   
  u^{n+1}_{ijk}   &\rightarrow& u^n_{ijk}- \frac{R_{ijk}}{\rd R_{ijk}},
\end{eqnarray}
where  $D_y^{p+}$ is  the  forward difference  operator  of order  $p$ in  the
$y$-direction,  $u_{ijk}$, $r_{ijk}$  and $y_{ijk}$ are the values of $u$, $r$,
and  $y$, respectively,  at  the lattice  location  $(i,j,k)$, and  $n$ is  an
iteration index. For example, in the case $p=2$ we obtain
\begin{eqnarray}
  &D_y^{2+} u_{ijk} = -\frac{3 u_{ijk} - 4 u_{i,j+1,k} 
    + u_{i,j+2,k}}{2 \Delta  y},& \label{eqn:Dy2p}\\  
  &\frac{\partial}{\partial u_{ijk} }D_y^{2+} u^n_{ijk} = -3 / 2\Delta y.& 
  \label{eqn:duDy2p}
\end{eqnarray}
Note that  Eq.~(\ref{eqn:ResBoundFD}) is linear  in $u_{ijk}$, so in  fact the
algorithm  is  equivalent  to  implementing  the  explicit  finite  difference
formula, with the advantage that its implementation is easier.
Since there are $p/2$ boundary buffer points, we have to specify a
method to obtain more than one buffer point. In our implementation the
method is stable if we update the values of the boundary points from
the inside to the outside of the domain. First, inside points are used
to get the first boundary point using the one-sided derivative. Then the
stencil is shifted by one from the inside to the outside, including
the first boundary point to compute data at the second boundary point,
and so forth.


\subsection{Results} 
\label{sec:results}

\subsubsection{Analytic test problems}
\label{subsubsec:analyticTest}

We  test \textsc{Olliptic}  with  three simple  elliptic  equations using  the
following procedure. Given the solution  $U^h$ on a mesh with grid-spacing $h$
and  an elliptic  operator $\DLie^h$,  we  calculate a  source $\rho^h$  which
satisfies the equation
\begin{equation}
\DLie^h U^h = \rho^h, \label{eqn:elliptic}
\end{equation}
and then we solve the equation to  obtain $u^h$ numerically. In this way it is
possible to calculate the error
\begin{equation} \label{eqn:NumErr}
E^h := \mid U^h-u^h \mid,
\end{equation}
where $\mid \cdot  \mid$ is a suitable norm.  We summarize  the grid setup for
our tests and puncture initial data in Table~\ref{Tab:GridSetup}.

\begin{table}[btp] 
  \begin{center}
    \caption{Grid setups used for tests and puncture initial data.  $l_r:=l-G$
      is  the number  of inner  refinement levels,  $L$ is  the length  of the
      numerical domain and $h_{min}$ is the grid size in the finest level (see
      text for details about each system).}\label{Tab:GridSetup}
    \begin{tabular}{l|ccc}
      \hline 
      \hline 
      System & Levels &  Length & Grid size \\ 
      & $l_r$ & $L$ & $h_{min}$ \\ 
      \hline 
           Test 1& 1-4& $4.8*2^{l-1}$& 1/20 \\
      Test 2 \& 3&   3&          20.0& $\{0.1,0.09,\ldots,0.02\}$\\ 
       1-puncture&   5&          40.0& $\{5/256,  1/64, 5/384\}$ \\ 
      2-punctures&   7&          40.0& $\{1/16, 1/32,  1/64\}$ \\ 
      3-punctures&   7&          50.0& 5/64 \\ 
      \hline 
      \hline
    \end{tabular}
  \end{center}
\end{table}

The  goals of the  first test  were to  estimate the  error introduced  by the
refinement method  and to  investigate the effectiveness  of the  algorithm to
solve non-linear equations. We have solved the equation
\begin{eqnarray} 
  \nabla^2  U(\vec{x})  +  U(\vec{x})^2  =  \rho_1(\vec{x})  \quad &
  \mathrm{for}&  \quad  \vec{x}  \in \Omega, \label{eqn:Test1Eq}\\  
  U(\vec{x}) = \epsilon  \, \e^{-\frac{1}{2}\vec{x}\cdot\vec{x}} \quad & 
  \mathrm{for}& \quad \vec{x} \in \partial \Omega, \label{eqn:Test1Sol}
\end{eqnarray}
where $\nabla^2$  is the three-dimensional  Laplace operator, and  $\Omega$ is
the  interior of  a  rectangular domain.   The  solution given  is a  Gaussian
function  with amplitude  $\epsilon=0.004$, in  this case  we use  a Dirichlet
boundary  condition.  We  have  solved the  equation  with a  single level  of
refinement in  a cube of length  $L=$4.8, and with  mesh size $dx=dy=dz=0.05$.
Using this solution as reference, we solve Eq.~(\ref{eqn:Test1Eq}), increasing
the number  of levels up  to 3 external  boxes. Due to the  Dirichlet boundary
condition the  numerical solution is  exact at the  boundary. We use  the norm
$L_\infty$ to calculate the relative error,
\begin{equation} \label{eqn:RelErr}
R := \frac{\mid U^h-u^h \mid}{\vert U^h \vert},
\end{equation}
and  as measurement  of  the error  introduced  by the  refinement method,  we
calculate  the difference  between the  error using  more than  one refinement
level and the reference solution, $\Delta R = \vert R(l>1)-R(l=0) \vert$.  The
results are summarized in Table~\ref{Tab:TestOll1}.

\begin{table}[btp] 
  \begin{center}
    \caption{Results of test 1, where $p$ is the order of the stencil which we
      use to solve  the equation, $l$ is the number  of refinement levels, $R$
      is the relative  error calculated in the finest level  and $\Delta R$ is
      the comparison with the reference solution. }\label{Tab:TestOll1}
    \begin{tabular}{crrrrrrrr}
      \hline   
      \hline  
      p&\multicolumn{2}{c}{2}  &\multicolumn{2}{c}{4} &\multicolumn{2}{c}{6}  
      &\multicolumn{2}{c}{8}\\  
      \hline 
      $l$&  $R$&  $\Delta R$&  $ R$&  $\Delta  R$& $  R$& $\Delta  R$& $  R$&
      $\Delta R$ \\
      &\multicolumn{2}{c}{($\times 10^{-4}$)} 
      &\multicolumn{2}{c}{($\times 10^{-7}$)}
      &\multicolumn{2}{c}{($\times 10^{-10}$)}
      &\multicolumn{2}{c}{($\times 10^{-12}$)}\\ 
      \hline  
      1&  3.83&    -&  4.36&     -&   9.29&      -&   3.05&      -\\ 
      2&  4.57& 0.74& 12.56&  8.20&  32.81&  23.52&  67.25&  64.20\\ 
      3&  5.23& 1.40& 15.42& 11.06& 105.37&  96.08& 207.75& 204.70\\ 
      4&  5.54& 1.71& 16.93& 12.56& 139.53& 130.24& 284.75& 281.70\\ 
      \hline 
      \hline
    \end{tabular}
  \end{center}
\end{table}

The results  for the non-linear  Eq.~(\ref{eqn:Test1Eq}) show that  using high
order  schemes  gives  a  significant  improvement  in  the  accuracy  of  the
solution. Increasing the order from $p$ to $p+2$ decreases $R$ by almost three
orders of magnitude.

In order to  test the implementation of the  Robin boundary condition,
we use a second trial function,
\begin{eqnarray} 
  \nabla^2 U(\vec{x}) = \rho_2(\vec{x}) \quad &\mathrm{for}& \quad 
  \vec{x} \in \Omega,\label{eqn:Test2Eq}\\ 
  U(\vec{x}) = \epsilon \frac{\tanh (r)}{r} \quad& 
  \mathrm{for}& \quad \vec{x} \in \partial \Omega, \label{eqn:Test2Sol}
\end{eqnarray}
where $r:=\vert\vec{x}\vert$.   The solution $U$  is a function which  has the
asymptotic behaviour  given by Eq.~(\ref{eq:DirichBC}) with  $A=0$, $B=1$, and
$q=1$. In this case  we look at the convergence of our  numerical data using 3
levels of refinement  in a cubic domain of length 20,  and using 9 resolutions
going  from  0.1  to  0.02  in  the finest  level.  For  a  finite  difference
implementation of order $p$, for $h \ll 1$, we expect
\begin{equation} \label{eqn:ErrOrd}
E^h \simeq C h^p,
\end{equation}
where $E^h$, is  given by Eq.~(\ref{eqn:NumErr}) using the  $L_2$ norm, $h$ is
the mesh size, and $C$ is constant with respect to $h$.  After calculating the
logarithm of Eq.~(\ref{eqn:ErrOrd}) we get a linear function of $p$,
\begin{equation} \label{eqn:ErrOrdLog}
\ln (E^h) \simeq p \ln(h) + C'.
\end{equation}
Using this expression with our data and doing a linear regression analysis, we
estimate the convergence order  $\mathcal{P}$ for our numerical experiment (in
the best case  $\mathcal{P} \longrightarrow p$ as $h  \longrightarrow 0$).  As
measurement of the error we use  the standard deviation and the coefficient of
variation of our data.  The results are displayed in Table~\ref{Tab:TestOll2}.
\begin{table}[btp] 
  \begin{center}
    \caption{Convergence test  for the Robin  boundary condition. Here  $p$ is
      the order  of the finite difference, $\mathcal{P}$,  $\sigma$, and $c_v$
      are the mean,  the standard deviation, and the  coefficient of variation
      of the  convergence order  for our numerical  experiments, respectively,
      and $\Delta \mathcal{P}$  is the relative deviation of  our results with
      respect to $p$.}\label{Tab:TestOll2}
    \begin{tabular}{crrrr}
      \hline 
      \hline 
      $p$ & $\mathcal{P}$  & $\sigma$ & $c_v$ & $ \Delta \mathcal{P} $\\  
      \hline 
      2& 2.002& 0.0002& 0.009\%& 0.10\%\\ 
      4& 3.994& 0.0005& 0.013\%& 0.15\%\\  
      6& 5.985& 0.0013& 0.022\%& 0.26\%\\ 
      8& 7.969& 0.0020& 0.026\%& 0.39\%\\ 
      \hline 
      \hline
    \end{tabular}
  \end{center}
\end{table}

We  have obtained  an accurate  implementation of  the boundaries  for problem
(\ref{eqn:Test2Eq})-(\ref{eqn:Test2Sol}),  where  the  difference between  the
theoretical convergence  order and  the experimental one  is less  than 0.5\%.
However,  note  that the  convergence  at  the  boundary depends  on  specific
properties of the test problem.

For  the last  analytic  test, we  verify the  accuracy  of the  method for  a
function which  is $C^\infty_{\vec{0}}:=C^\infty(\R^3 \setminus \{\vec{0}\})$.
The problem to solve was
\begin{eqnarray} 
  \nabla^2 U(\vec{x}) = \rho_3(\vec{x}) \quad &\mathrm{for}& \quad   
  \vec{x} \in \Omega,\label{eqn:Test3Eq}\\  
  U(\vec{x}) = r^k \quad& \mathrm{for}&  \quad 
  \vec{x} \in \partial \Omega, \label{eqn:Test3Sol}
\end{eqnarray}
where we  set $k=3$  or $k=5$, $r:=\vert\vec{x}\vert$,  and $U$  is $C^\infty$
everywhere except at the origin, where it is $C^{k-1}$.
\begin{table}[btp] 
  \begin{center}
    \caption{Convergence for  a solution which  is $C^\infty_{\vec{0}}$.  Here
      $k$ is the exponent given in (\ref{eqn:Test3Sol}).}\label{Tab:TestOll3}
    \begin{tabular}{crrrr|rrrr}
      \hline
      \hline
      k& \multicolumn{4}{c}{3}& \multicolumn{4}{c}{5}\\ 
      \hline 
      $p$& 
      $\mathcal{P}$& $\sigma$& $c_v$& $\Delta \mathcal{P}$&  
      $\mathcal{P}$& $\sigma$& $c_v$& $\Delta \mathcal{P}$\\ 
      \hline  
      2& 2.003& 0.0003& 0.013\%&  0.16\%& 1.999& 0.0001& 0.005\%&  0.06\%\\  
      4& 3.782& 0.0088& 0.233\%&  5.46\%& 3.995& 0.0003& 0.007\%&  0.12\%\\ 
      6& 3.848& 0.0067& 0.175\%& 35.86\%& 5.715& 0.0124& 0.216\%&  4.74\%\\ 
      8& 3.836& 0.0038& 0.098\%& 52.05\%& 5.868& 0.0294& 0.500\%& 26.65\%\\ 
      \hline 
      \hline
    \end{tabular}
  \end{center}
\end{table}

We use  the procedure of  the second test  to estimate the  convergence order,
changing the  equation and  the boundaries  (in this case  we use  a Dirichlet
boundary  condition). The  result of  our numerical  experiments  (detailed in
Table~\ref{Tab:TestOll3}) shows that  the overall convergence of the numerical
solution calculated using a  standard finite differencing scheme is restricted
by  the differentiability of  the analytical  solution. The  convergence order
close to  the origin (within a  few grid points) is  the same as  the order of
differentiability and improves significantly  moving away from the origin. The
accuracy of the solution behaves in the same manner.


\subsubsection{Single puncture initial data}
\label{subsubsec:1punctureID}

\begin{figure}[btp]
  \centering 
  \includegraphics[scale=0.7]{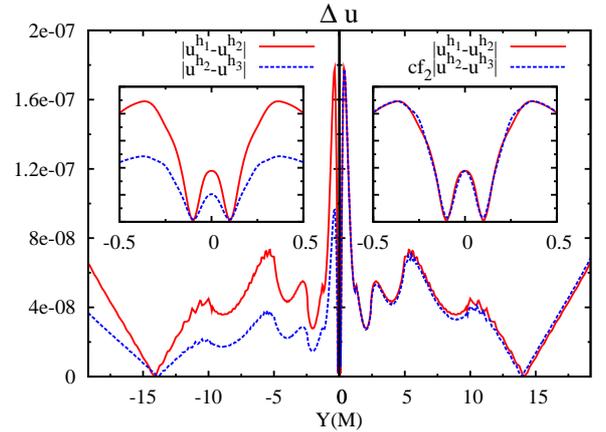}
  \caption{Regular part $u$ of the conformal factor along the $Y$-axis
    of a single puncture with vanishing spin parameter and with linear
    momentum $P_y=0.2M$. Shown is a convergence test without scaling
    (left) and with scaling (right) for second-order convergence using
    $\mathrm{cf}_2=1.8409$.}
  \label{fig:SBHpy0.2O2_ECY} 
\end{figure}

After  calibrating our  code, we  calculate the  Hamiltonian constraint  for a
single puncture. We tested  the convergence of our second-order implementation
for a single  boosted puncture ($P^i=0.2 \, \delta_2^iM$,  $S^i=0$) by looking
at  the value  of  the regular  part $u$  of  the conformal  factor along  the
$Y$-axis for  a cubic domain  of length $40M$,  5 levels of refinement,  and 3
resolutions $h_1=(5/8)M$, $h_2=4h_1/5$, and  $h_3=2h_1/3$ in the coarse level.
In  Fig.~\ref{fig:SBHpy0.2O2_ECY},  we show  rescaled  and  unscaled data  for
positive and negative values of $Y$, respectively.

We plot  the values  of $\vert  u^{h_1} - u^{h_2}\vert$  and  $\vert  u^{h_2} -
u^{h_3} \vert$ for $Y<0$  on the left, and on the right  values for $Y>0$ with
$\vert u^{h_2} - u^{h_3}  \vert$ multiplied by a factor  $\mathrm{cf}_2=1.8409$
which corresponds  to the  proper scaling of  second order.  The lines  in the
right panel  of the plot  coincide almost everywhere, indicating  second order
convergence.  We  also show details of a  region close to the  puncture in the
insets.

\begin{figure}[btp]
  \centering 
  \includegraphics[scale=0.7]{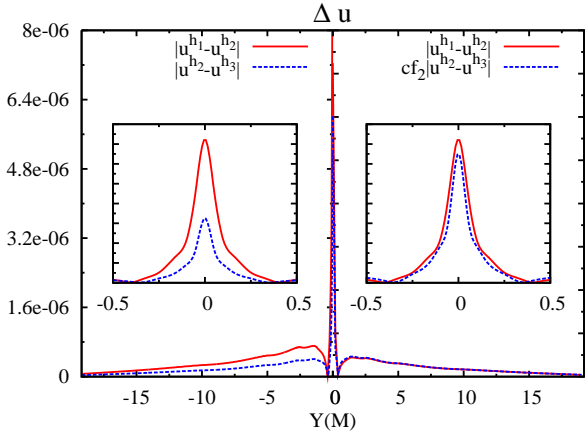}
  \caption{Regular part  $u$ of the conformal  factor along the  $Y$-axis of a
    single   puncture   with  vanishing   linear   momentum   and  with   spin
    $S_y=0.2M$. Shown is a convergence test without scaling
    (left) and with scaling (right) for second-order convergence using
    $\mathrm{cf}_2=1.8409$.}
  \label{fig:SBHsy0.2O2_ECY} 
\end{figure}

We  perform  a similar  test  calculating  spinning  black hole  initial  data
($P^i=0$, $S^i=0.2 \,  \delta_2^iM$).  Fig.~\ref{fig:SBHsy0.2O2_ECY} shows the
result  of the  convergence test  for this  case where  we found  second order
convergence again.

\begin{figure}[btp]
  \centering 
  \includegraphics[scale=0.7]{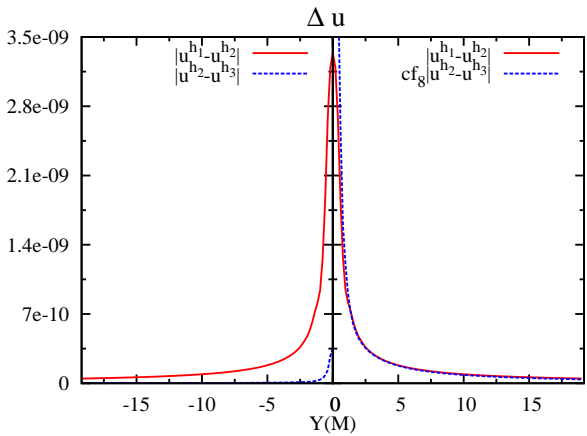}
  \caption{Regular part  $u$ of the conformal  factor along the  $Y$-axis of a
    single  puncture with vanishing  spin parameter  and with  linear momentum
    $P_y=0.2M$. Eighth-order  convergence   
    of   $u$ is obtained far from the puncture ($\mathrm{cf}_8=6.4637$).}
  \label{fig:SBHpy0.2O8_ECY} 
\end{figure}

As an  example of a high  order solution, in  Fig.~\ref{fig:SBHpy0.2O8_ECY} we
show the convergence test for the eighth order scheme of the boosted puncture.
In this  case the  plot shows  a drop of  the convergence  ratio close  to the
puncture. However, far from the puncture the convergence behavior is better.

\begin{figure}[btp]
  \centering 
  \includegraphics[scale=0.7]{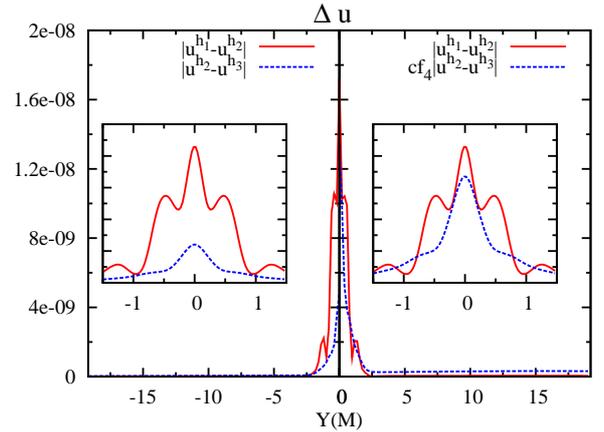}
  \caption{Regular part $u$ of the conformal factor along the $Y$-axis
    of a single puncture with vanishing linear momentum and with spin
    $S_y=0.2M$. Convergence test without scaling (left) and
    with scaling (right) for fourth-order convergence using
    $\mathrm{cf}_4=2.7840$.}
  \label{fig:SBHsy0.2O4_ECY} 
\end{figure}

In Fig.~\ref{fig:SBHsy0.2O4_ECY}  we plot the  results for the  spinning black
hole,  obtained by  using our  fourth order  implementation.  Compared  to the
boosted puncture,  in this case we  see better behavior close  to the puncture
(the solution  of the 8th  order spinning puncture  is similar to  the boosted
case). Far   from  the  puncture  the  convergence   ratio  is  approximately
second order.

As we  saw in  our third  test and in  our numerical  experiment for  a single
boosted or  spinning puncture, the convergence  rate of the  high order finite
differencing   scheme  for  functions   $C^\infty_{\vec{0}}$  drops   near  to
$\vec{0}$.  This is  a well  known property  of high  order  finite difference
schemes (e.g.~\cite{Lin01,BaiBra87}). We review  some basics of this effect in
Appendix~\ref{appendixA}.  Nevertheless, as  we show in \ref{sec:results2} and
in the two-punctures test (see  below), the numerical solution produced by our
high order  implementation seems  to be accurate  enough to  perform numerical
evolutions of  multiple black holes. The  errors close to the  puncture do not
modify significantly the convergence during the evolution.


\subsubsection{Two-puncture initial data}
\label{subsubsec:2punctureID}

As a  test for  a binary  system we set  the parameters  for two  punctures to
$x_1=-x_2=3M$,  $P^i_1=-P^i_2=0.2  \,  \delta^i_2M$.  This  configuration  was
studied before  using a  single-domain spectral method  \cite{AnsBruTic04}. We
compared the result of our new code with the solution produced by the spectral
solver.   For  the  spectral  solution  we use  $n_A=n_B=40$  and  $n_\phi=20$
collocation-points (see reference for details about the definition of spectral
coordinates  $(A,B,\phi)$).  We calculate  the multigrid  solution in  a cubic
domain  of  length  $40M$,  7  levels  of  refinement  and  3  resolutions  of
$h_1=(1/16)M$, $h_2=h_1/2$ and $h_3=h_1/4$ in the finest level.
  
\begin{figure}[btp]
  \centering 
  \includegraphics[scale=0.7]{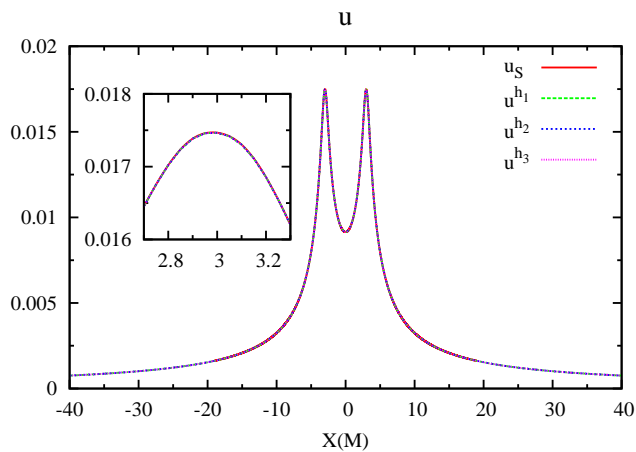}
  \caption{Comparison between  the   numerical  solution   of  the
    Hamiltonian  constraint calculated using  a single-domain  spectral method
    and the  high-order multigrid  solver.   The  plot  shows $u$  along  the
    $X$-axis produced by the spectral  code (denoted by $u_S$) and using three
    resolutions  calculated  with  the  eighth  order  implementation  of  the
    multigrid code (labels $u^{h_i}$). }
  \label{fig:ABTO8Xfull} 
\end{figure}

Fig.~\ref{fig:ABTO8Xfull} is  a plot similar to  Fig.~5 of \cite{AnsBruTic04}.
We compare  the spectral  solution with the  eighth order  multigrid solution.
The fact that the four lines coincide  on the scale of the plot (3 resolutions
of multigird and  one spectral solution) indicates that  the two methods agree
with each other on the whole domain.

\begin{figure}[btp]
  \centering
  \includegraphics[scale=0.7]{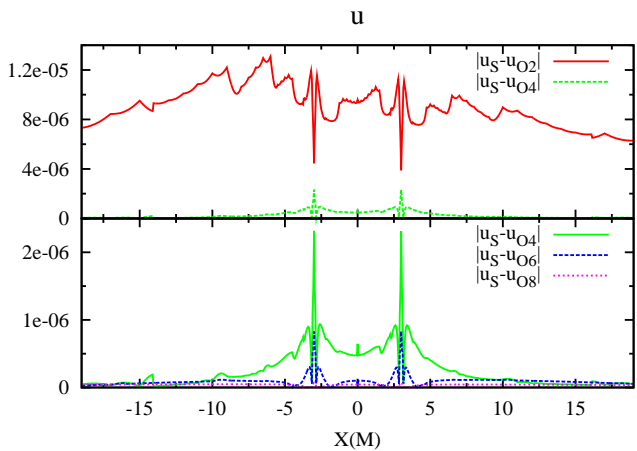}
  \caption{Absolute value of the differences between the numerical
    solution of $u$ for the second, fourth, sixth, and eighth order
    finite difference implementation and the spectral solution. The
    upper part shows the result for second and fourth order, and the
    bottom part for fourth, sixth, and eighth order.}
  \label{fig:ABTO2O4O6O8Spec} 
\end{figure}

Using the  same setting we solve  the Hamiltonian constraint  with the second,
fourth, and sixth order stencil of  the multigrid code. Then we use the highly
accurate  solution of  the  spectral code  as  reference to  compare with  the
different orders.  As we showed before in  the case of a  single puncture, the
accuracy close  to the  puncture decreases. However,  the comparison  with the
spectral  code (see  Fig.~\ref{fig:ABTO2O4O6O8Spec}) shows  that using  high
order finite differencing stencils improves the accuracy of the solution.

Spectral  methods  produce in  general  more  accurate  solutions to  elliptic
equations  than  those  obtained  by finite  difference  methods~\cite{boy01}.
However, in order to take full advantage of the spectral method for punctures,
it is necessary to construct a special set of coordinates. Indeed, there exist
coordinates in  which the conformal correction  $u$ is smooth  at the puncture
\cite{AnsBruTic04}. Although these coordinates are in principal applicable for
both  spectral  and  finite  differencing  methods, the  resulting  grids  are
specific  to two  black holes.  Generalizing that  approach to  more  than two
punctures is an interesting but non-trivial challenge that we do not pursue in
this work.

Using  finite difference  multigrid  methods with  Cartesian coordinates,  one
advantage  of the  puncture construction  is that  it is  possible  to produce
accurate solutions of the Hamiltonian constraint for multiple black holes with
minimal changes to a code prepared for binaries.


\subsubsection{Three-puncture initial data}
\label{subsubsec:3punctureID}

In  previous   work  on   the  numerical  evolution   of  three   black  holes
\cite{CamLouZlo07f,LouZlo07a}, the  Hamiltonian constraint has  been specified
using            an            approximate            solution            (see
\cite{LouZlo07a,Lag03a,DenBauPfe06,GleKhaPul99,GleNicPri97}).   We compare our
numerical  solution with  the approximate  solution (which  we  implemented as
well)  for  the  set  of  parameters  labeled  3BH102  given  in  Table~I  of
\cite{CamLouZlo07f}, see our Table~\ref{Tab:IDpar1}.
\begin{figure}[btp]
  \centering
  \includegraphics[scale=0.7]{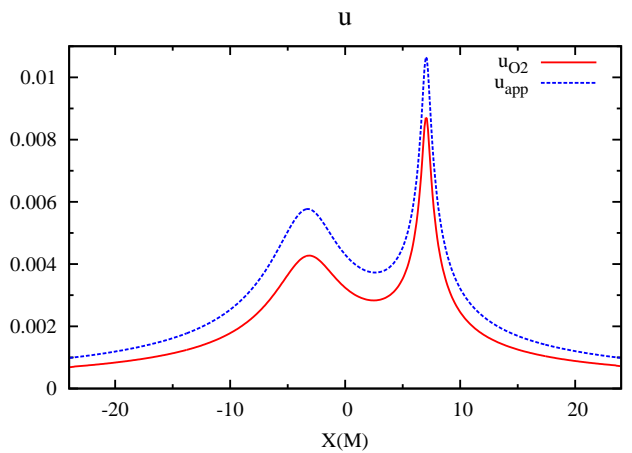}
  \caption{Plot  of  $u$  along  $X$-axis  for system  3BH102,  comparing  the
    approximate solution  with a second  order numerical solution.  The higher
    order numerical  solutions would not be distinguishable  from second order
    in this plot, compare Fig.~\ref{fig:ABTO2O4O6O8Spec}.}
  \label{fig:TBH102App_vs_O2} 
\end{figure}

In  Fig.~\ref{fig:TBH102App_vs_O2} we  show a  plot of  the  solution obtained
using a cubic domain of length $50M$, a mesh size $h=0.5M$ in the coarse level
and 9  levels of  refinement. The approximate  solution was calculated  in the
same numerical grid. The result shows a significant difference between the two
methods, and, as we will show  later in \ref{sec:results2}, that fact leads to
a quantitative and qualitative difference for evolutions.


\section{NUMERICAL EVOLUTION OF THREE BLACK HOLES}
\label{sec:num_evol_3bhs}

In the mid  1960's, Hahn and Lindquist started  the numerical investigation of
colliding  black holes  \cite{HahLin64}. After  more  than forty  years and  a
series         of        breakthroughs         starting         in        2005
\cite{pre05,CamLouMar05,BakCenCho05b,BruTicJan03,SchPfeLin06},  the  numerical
relativity  community  is now  able  to  produce  stable black  hole  inspiral
simulations  and  to compute  gravitational  waves  signals.  The most  common
formulation used to  perform numerical evolutions of black  holes is based on
the work of  Shibata and Nakamura \cite{ShiNak95a}, and  Baumgarte and Shapiro
\cite{BauSha98b} and is known as the BSSN formulation.


\subsection{Techniques}
\label{sec:techniques}

We have performed the three black hole simulations using the \textsc{BAM} code
as    described    in    \cite{BruGonHan06,BruTicJan03},    and    with    the
\textsc{AMSS-NCKU}  code \cite{CaoYoYu08}.   In  \textsc{BAM} we  use a  sixth
order discretization for the spatial derivatives \cite{HusGonHan07} and fourth
order  accurate  integration  in  time.   Initial data  are  provided  by  the
\textsc{Olliptic} code.  Gravitational waves are calculated in the form of the
Newman-Penrose  scalar  $\Psi_4$  according  to  the  procedure  described  in
Sec.~III  of \cite{BruGonHan06}.   We use  the BSSN  system together  with the
$1+\log$        and       gamma        freezing        coordinate       gauges
\cite{AlcBru00,BakBruCam01,AlcBruDie02}  as  described  in  \cite{BruGonHan06}
(choosing in particular  the parameter $\eta=2/M$ in the  gamma freezing shift
condition).   All    the   runs   are   carried   out    with   the   symmetry
$(x,y,z)\rightarrow(x,y,-z)$ in  order to reduce the  computational cost.  The
Courant  factor,  $\mathcal{C}:=\Delta  t/h_i$,   seems  to  be  an  important
ingredient to reach convergence. For long evolutions (evolution time $t>200$),
we set $\mathcal{C}=1/4$, in other cases we use $\mathcal{C}=1/2$.

The \textsc{AMSS-NCKU}  code is an extended  version of the  code described in
\cite{CaoYoYu08}.  Instead of  \textsc{G}r\textsc{ACE}, we constructed our own
driver combining \textsc{C++} and  \textsc{Fortran 90} to implement moving box
style  mesh  refinement.  Regarding  the  numerical scheme  dealing  with  the
interface  of neighbor  levels, we  closely  follow the  methods described  in
\cite{BruGonHan06,YamShiTan08}.  \textsc{AMSS-NCKU}  can implement both  the 6
point  buffer  zone  method   \cite{BruGonHan06}  and  interpolation  at  each
sub-Rung-Kutta  step  \cite{YamShiTan08}.  Our  tests  show little  difference
between these two methods.  For simplicity, all simulations presented here use
the 6  point buffer zone  method.  In order  to do 3rd order  interpolation in
time, we  need three time levels of  data.  At the beginning  of the numerical
evolution, we  use a 4th  order Rung-Kutta method  to evolve the  initial data
backward one  step to get  the data on  time level $t=t_0-\Delta t$  for every
mesh level \cite{SchHawHaw03}. Here $\Delta t$ is different for different mesh
levels.   We have reproduced  the results  published in  \cite{CaoYoYu08} with
this driver.  All  of the tests involved fixed mesh  refinement and agree very
well  with  the results  obtained  with  \textsc{G}r\textsc{ACE}.   As to  the
Einstein equation solver, we replaced  the ICN method used in \cite{CaoYoYu08}
by  a 4th  order Runge-Kutta  method.   The Sommerfeld  boundary condition  is
implemented with 5th order interpolation.


\subsection{Results}
\label{sec:results2}

With  the  \textsc{BAM}  code  we  simulate three  black  holes  with  initial
parameters as  given in Table~\ref{Tab:IDpar1}.  In the  first experiments, we
focus  on runs  that use  the  initial data  parameters of  runs ``3BH1''  and
``3BH102'' in \cite{CamLouZlo07f}. We evolve this data with both the numerical
initial data and the approximate  solution to the conformal factor. We compare
the puncture  tracks and the extracted  wave forms with those  produced by the
\textsc{AMSS-NCKU} code. The puncture tracks  give a convenient measure of the
black hole  motion. It is much  more cumbersome to compute  the event horizon,
which we do for a simple black hole triple in \cite{ThiBruGal10}.  Finally, we
discuss some  results for the  evolution of four  black holes which  show some
additional properties of multiple black holes evolutions.
\begin{table}[btp] 
  \begin{center}
    \caption{Initial data parameters}\label{Tab:IDpar1}
    \begin{tabular}{lrrrr}
      \hline 
      \hline 
      Parameter & 3BH1 & 3BH102 & FBHSR & FBH3w\\ 
      \hline
      $x_1/M$  & -2.4085600 & -3.5223800 &-10.0000 & -10.000000\\  
      $y_1/M$  &  2.2341300 &  2.5850900 &  0.0000 &  0.0000000\\  
      $p^x_1/M$& -0.0460284 &  0.0782693 & -0.0100 &  0.0000000\\
      $p^y_1/M$& -0.0126181 & -0.0433529 &  0.0750 &  0.0524275\\ 
      $m_1/M$  &  0.3152690 &  0.3175780 &  0.2500 &  0.2500000\\  
      $x_2/M$  & -2.4085600 & -3.5246200 & -6.0000 & -6.0000000\\ 
      $y_2/M$  & -2.1053400 & -2.5850900 &  0.0000 &  0.0000000\\ 
      $p^x_2/M$&  0.1307260 & -0.0782693 &  0.0000 &  0.0000000\\ 
      $p^y_2/M$& -0.0126181 & -0.0433529 & -0.0750 & -0.0524275\\  
      $m_2/M$  &  0.3152690 &  0.3175780 &  0.2500 &  0.2500000\\  
      $x_3/M$  &  4.8735000 &  7.0447600 &  6.0000 &  8.0000000\\  
      $y_3/M$  &  0.0643941 &  0.0000000 &  0.0000 &  0.0000000\\  
      $p^x_3/M$& -0.0846974 &  0.0000000 &  0.0000 &  0.0000000\\  
      $p^y_3/M$&  0.0252361 &  0.0867057 &  0.0750 & -0.0781250\\  
      $m_3/M$  &  0.3152690 &  0.3185850 &  0.2500 &  0.2500000\\ 
      $x_4/M$  &  -         & -          & 10.0000 & 10.0000000\\  
      $y_4/M$  &  -         & -          &  0.0000 &  0.0000000\\  
      $p^x_4/M$&  -         & -          &  0.0100 &  0.0000000\\
      $p^y_4/M$&  -         & -          & -0.0750 &  0.0781250\\ 
      $m_4/M$  &  -         & -          &  0.2500 &  0.2500000\\  
      \hline 
      \hline
    \end{tabular}
  \end{center}
\end{table}

\subsubsection{Three black holes}
\label{sec:3BHs}

\begin{figure}[btp]
  \centering
  \includegraphics[scale=0.7]{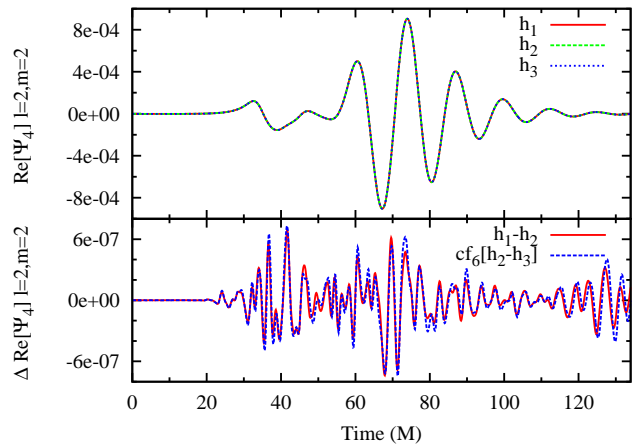}
  \caption{Real part of $\Psi_4$ (mode $l=m=2$) calculated at $r=40M$
    for system 3BH1. The lower panel shows the convergence test for
    6th order ($\mathrm{cf}_6=1.9542$).}
  \label{fig:TBH1_rpsi4_3resConv_OllipticID_6th}
\end{figure}
System 3BH1 is  a short simulation which is useful  for convergence tests.  We
use our sixth  order implementation to calculate initial  data, a cubic domain
of   length  $1052M$,   10   levels  of   refinement   and  three   resolution
$h_1=(125/12)M$, $h_2=6h_1/7$  and $h_3=2h_1/3$ on the finer  level.  We have
obtained roughly  sixth order convergence  for the gravitational  waveform, as
shown in  Fig.~\ref{fig:TBH1_rpsi4_3resConv_OllipticID_6th}.  Our results show
a $\Psi_4$ waveform similar to  that shown in Fig.~16 of \cite{LouZlo07a}. For
this  evolution,  we  did  not   find  a  significant  difference  when  using
approximate  initial data  or  solving the  constraint equations  numerically.
However,  the accuracy  of the  numerical initial  data is  important  for the
numerical  result to converge.  Our first  test using  \textsc{BAM}'s elliptic
solver (which  is a second-order  multigrid solver) showed that  for long-time
simulation it  is important to  improve the accuracy  of the initial  data for
multiple black hole evolutions.

\begin{figure}[btp]
  \centering
  \includegraphics[scale=0.7]{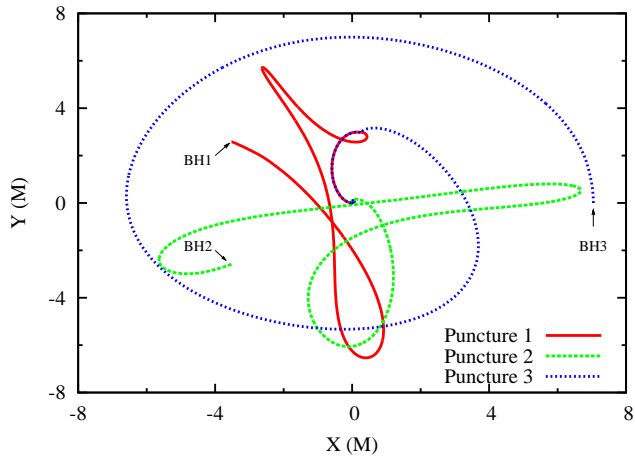}
  \caption{
    Puncture tracks for system 3BH102 using approximate initial data.
    (see  text, Sec.~\ref{subsubsec:3punctureID}).}
  \label{fig:TBH102_puncturetrack_AppID}
\end{figure}

Our second example is black hole configuration 3BH102, which we consider first
for approximate initial data, and  later for the numerical solution.  This set
of parameters is a system which, starting with approximate initial data, leads
to trajectories forming  a nice figure similar to  the Greek letters $\gamma$,
$\sigma$  and $\tau$ (see  Fig.~\ref{fig:TBH102_puncturetrack_AppID}, computed
with \textsc{BAM}).   Our convergence test  for this system  shows sixth-order
(see Fig.~\ref{fig:TBH102_rpsi4_3resConv}), with  small deviations from second
and  fourth order  which are  consistent with  the accuracy  of  the evolution
method of our code.

\begin{figure}[btp]
  \centering
  \includegraphics[scale=0.7]{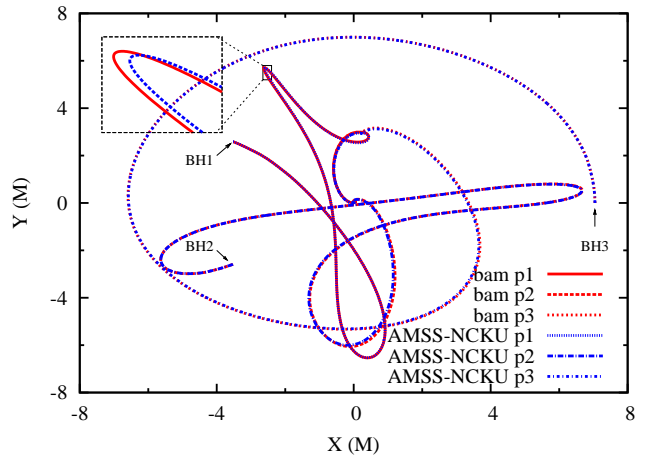}
  \caption{
    Puncture tracks for system 3BH102 using approximate initial data comparing
    results for the \textsc{BAM} and \textsc{AMSS-NCKU} codes.  The difference
    in  the  trajectories is  small,  and the  results  agree  in the  general
    shape.  Note  that  AMSS-NCKU  uses  fourth-order  spatial  discretization
    instead of sixth-order which is implemented in \textsc{BAM}.}
    \label{fig:TBH102_puncture_App_bamAMSS} 
\end{figure}

\begin{figure}[btp]
  \centering
  \includegraphics[scale=0.7]{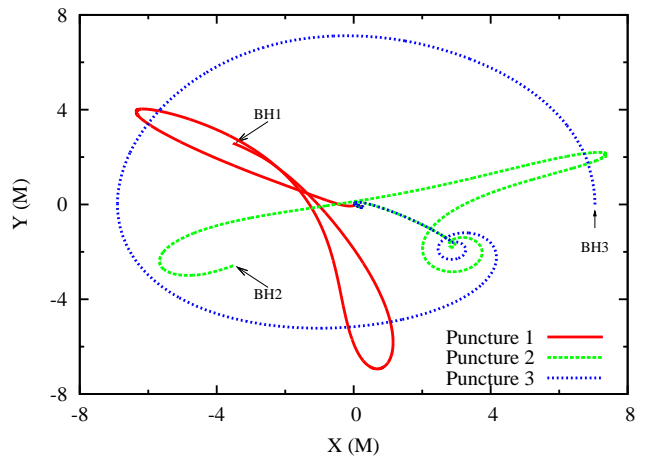}
  \caption{ Puncture tracks for system  3BH102 using the numerical solution of
    the Hamiltonian constraint with the 8th order multigrid method. There is a
    drastic change  in the  puncture tracks compared  to the evolution  of the
    approximate  initial  data, in  particular  the  black  holes merge  in  a
    different order (compare Fig.~\ref{fig:TBH102_puncturetrack_AppID}).}
  \label{TBH102_puncturetrack_OllipticID_6th}
\end{figure}

Comparing with Fig.~3  of \cite{CamLouZlo07f}, there is a  small but noticable
difference in  the puncture tracks  of roughly up  to $1M$ in  the coordinates
compared to  our results.   There are several  possible explanations  for this
difference. Evolutions of multiple black  holes are sensitive to small changes
in the grid setup and initial data.  We tested possible sources of errors, for
example introduced  by numerical  dissipation or finite  resolution.  Changing
these lead to negligible changes in  the trajectories on the scale of the plot
and  do  not seem  to  explain the  existing  difference.  However, since  the
deviation from  \cite{CamLouZlo07f} does not  change the qualitative  shape of
the tracks, we conclude that we have consistently reproduced that simulation.

Alternatively,  we  can compare  the  paths  of  the punctures  obtained  with
\textsc{BAM}  with  those  produced   by  the  \textsc{AMSS-NCKU}  code.   The
implementation of the approximate initial  data was done independently for the
two codes, and in both cases the formula from \cite{LouZlo07a} is used. We see
in  Fig.~\ref{fig:TBH102_puncture_App_bamAMSS} that the  results from  the two
codes  agree  within  a  maximum  difference  of about  $0.2M$  in  the  given
coordinates, or 2\% with respect to an orbital scale of $10M$.  An analysis of
the $l=m=2$ mode of $\Psi_4$ showed that there are differences in the phase of
about 0.4\% and of about 2\% in the amplitude.

When comparing  codes, recall that  the \textsc{BAM} evolutions use  6th order
spatial  differencing, \textsc{AMSS-NCKU}  4th  order, and~\cite{CamLouZlo07f}
also 4th order for the figures,  pointing out that there was little difference
to an 8th order run. Our  conclusion is that differences due to resolution are
small,  and they  are significantly  smaller  than the  changes introduced  by
replacing  the  approximate  initial  data  by a  numerical  solution  of  the
Hamiltonian constraint.
 
\begin{figure}[btp]
  \centering
  \includegraphics[scale=0.7]{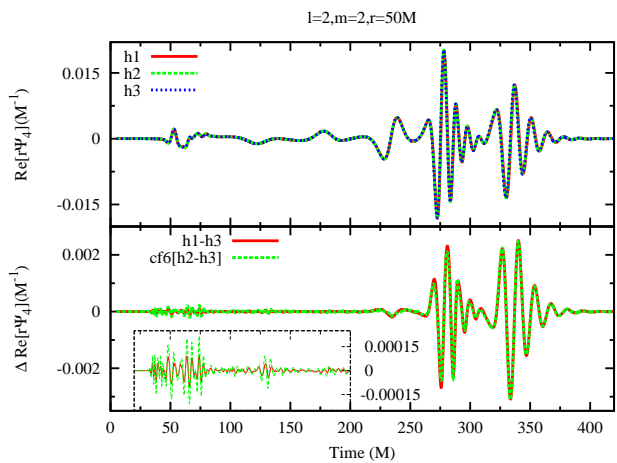}
  \caption{Real part  of $r \Psi_4$  (mode $l=m=2$) calculated at  $r=50M$ for
    system 3BH102 using  approximate initial data.  The upper  panel shows the
    $r \Psi_4$ waveform  for 3 resolutions, the bottom  plot shows sixth-order
    convergence scaling with a factor $\mathrm{cf}_6=1.9542$. }
  \label{fig:TBH102_rpsi4_3resConv}
\end{figure}

\begin{figure}[btp]
  \centering
  \includegraphics[scale=0.7]{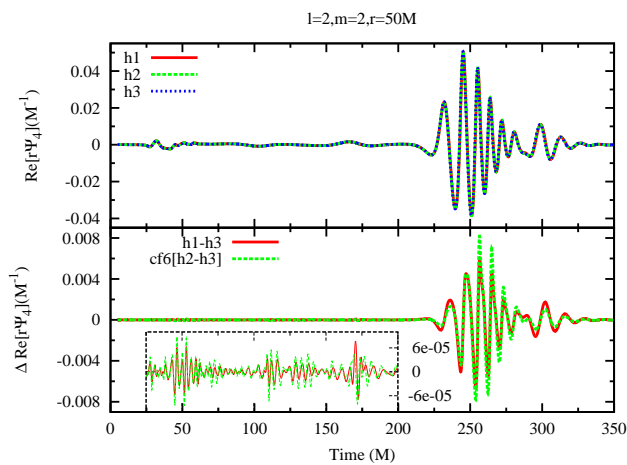}
  \caption{Real part  of $r \Psi_4$  (mode $l=m=2$) calculated at  $r=50M$ for
    system   3BH102  using   the   numerical  solution   of  the   Hamiltonian
    constraint.  The  upper  panel  shows   the  $r  \Psi_4$  waveform  for  3
    resolutions, the bottom plot  shows sixth-order convergence scaling with a
    factor  $\mathrm{cf}_6=1.9542$.   Near  the  first  merger  the  order  of
    convergence is closer to fourth-order.}
  \label{fig:TBH102_rpsi4_3resConv_OllipticID_6th}
\end{figure}

We  now  focus on  the  evolution of  system  3BH102  solving the  Hamiltonian
constraint with  the eighth-order multigrid  method. The $\Psi_4$  waveform and
convergence are  shown in Fig.~\ref{fig:TBH102_rpsi4_3resConv_OllipticID_6th}.
Note that we see approximately  sixth-order convergence in the waveform except
for the first merger where the convergence is close to 4th order.

As shown in Sec.~\ref{subsubsec:3punctureID},  for system 3BH102 the numerical
solution  of   the  Hamiltonian   constraint  differs  from   the  approximate
prescription.  As a consequence, the trajectories and waveform change. We show
the    paths    followed    by    the    punctures   for    this    case    in
Fig.~\ref{TBH102_puncturetrack_OllipticID_6th}.
Instead of the grazing collisions of  the previous evolution, in this case the
black  holes with labels 2 and  3 merge  after a  small inspiral,  producing a
higher amplitude in the wave. The second merger is almost a head-on collision,
which generates a smaller amplitude in  the wave. Notice that the order in 
which the black holes merge differs from the previous evolution.

Looking   at   the   wave    forms,   for   the   approximate   initial   data
Fig.~\ref{fig:TBH102_rpsi4_3resConv}  shows   a  relatively  large   burst  of
``junk''-radiation which does not converge. Solving the Hamiltonian constraint
we       see       a        better       convergence       behavior,       see
Fig.~\ref{fig:TBH102_rpsi4_3resConv_OllipticID_6th}.  Moreover, the difference
in the  junk-radiation between resolutions using the  approximate initial data
is  one order  of magnitude  bigger  than solving  the Hamiltonian  constraint
numerically (compare  the insets in  Figs.~\ref{fig:TBH102_rpsi4_3resConv} and
\ref{fig:TBH102_rpsi4_3resConv_OllipticID_6th}).

In the case  of a binary system  it is possible to produce  the same evolution
for   numerical  and   approximate  initial   data  by   adjusting   the  mass
parameter~\cite{LouZlo07a}.  In the case of  three black holes, there does not
seem  to be  a simple  procedure to  fit the  initial parameters  in  order to
reproduce  the same  trajectory  with both  types  of initial  data. We  tried
changes in the momentum, the mass, and the momentum and mass together, looking
at the  maximum of the regular  part $u$ of  the conformal factor in  order to
reduce the  difference between the  analytical prescription and  the numerical
data. The result  is not satisfactory, i.e.\  we did not find a  way to change
the parameters of  the approximate data to better  approximate the solution of
the Hamiltonian constraint,  and the large differences in  the puncture tracks
could not be removed.

\begin{figure}[btp]
  \centering
  \includegraphics[scale=0.7]{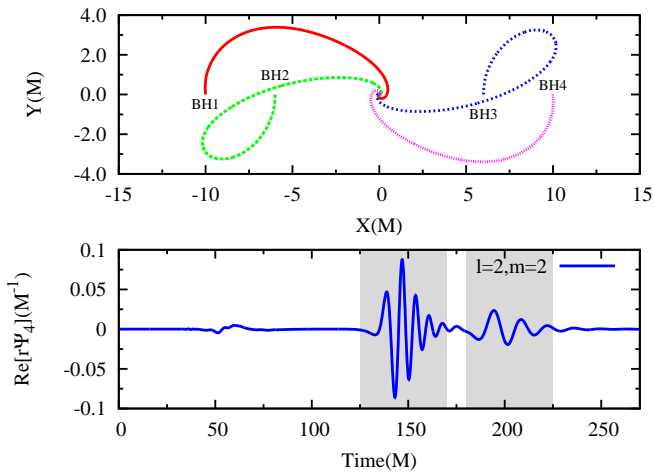}
  \caption{Symmetric configuration  of four black holes.   Top: Paths followed
    by the punctures with a rotational symmetry of 180$^o$ with respect to the
    $Z$-axis. Bottom: Real part of  the $l=m=2$ mode of $r\Psi_4$ extracted at
    $r=50M$, the  waveform shows two  mergers (around $t=150M$  and $t=200M$),
    first  between the  black holes labeled  BH1 and  BH4, and  the second
    merger between the resulting black hole and the black holes BH2 and BH3.}
  \label{fig:FBHd_puncturetrack_waves}
\end{figure}

\subsubsection{Four black holes}
\label{sec:4BHs}

Evolution of more than three black  holes is possible using the same approach.
We performed  several tests  for evolutions of  multiple black holes.  Here we
present two particular cases using four black holes.

The first  case is the system  that we call  FBHSR (see Table~\ref{Tab:IDpar1}
for  details about  the initial  parameters).  We  start with  four equal-mass
black holes  aligned on the $X$-axis,  and with an arbitrary  selection of the
initial momentum  symmetric with respect  to the $YZ$-plane.  The  black holes
follow         a         rotationally       symmetric        path         (see
Fig.~\ref{fig:FBHd_puncturetrack_waves}). From the waveform we can distinguish
two mergers. The first merger is  between the black holes labeled BH1 and BH4,
the black  hole generated by  that merger stays  in the origin until  a triple
merger occurs with the black holes BH2 and BH3.  The second merger is almost a
triple head-on collision with a small amplitude wave form.

\begin{figure}[btp]
  \centering
  \includegraphics[scale=0.7]{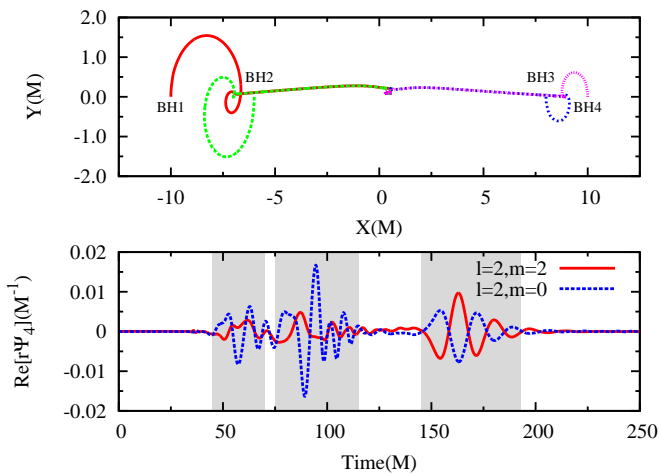}
  \caption{Triple  merger of  four black  holes.  Top:  Paths followed  by the
    punctures.  Bottom:  Real part  of the $l=m=2$  and $l=2$, $m=0$  modes of
    $r\Psi_4$ extracted  at $r=50M$. The waveform shows  three mergers (around
    $t=55M$,  $t=95M$, and  $t=170M$),  which are  more  easily identified  by
    looking at the mode $l=2$, $m=0$.}
  \label{fig:FBH3wc_puncturetrack_waves}
\end{figure}

The second case  (FBH3w in Table~\ref{Tab:IDpar1}) consists of  a quick merger
of  two binary  systems (see  Fig.~\ref{fig:FBH3wc_puncturetrack_waves}).  The
remaining black  holes merge  in an almost  head-on collision. Looking  at the
waveforms ($l=m=2$ and  $l=2$, $m=0$ modes of $\mathrm{Re}(r  \Psi_4)$) we can
identify the three collisions, the  second with higher amplitude.  The initial
parameters were  chosen by trial  and error in  order to produce this  kind of
waveform.  With  a larger  separation and more  careful choice of  the initial
parameters it would be possible to find a stronger and more distinctive merger
signal. However, the above examples are only intended to be an illustration of
the kind of waveforms that can be generated by mergers of four black holes.

\section{DISCUSSION}
\label{sec:discussion}

We have presented a numerically elliptic solver, \textsc{Olliptic}. As a first
application, we  solve the Hamiltonian constraint to  obtain numerical initial
data  for  multiple  black  hole evolutions.  \textsc{Olliptic}  implements  a
high-order multigrid method,  which is parallelized and uses  a box-based mesh
refinement. The tests  and first applications of the code  showed that the new
code seems  to be  sufficiently accurate for  our purposes. However,  we found
that close  to the puncture  the convergence rate  is less than  that desired,
which is expected  for puncture data (see Appendix  \ref{appendixA}). The drop
in the convergence close to the  punctures is not reflected in the convergence
of the  evolution. Nevertheless,  we are considering  to modify  the numerical
scheme in order to improve the accuracy close to the puncture.

We have  shown evolutions of three and  four black holes which  use as initial
data solutions to  the Hamiltonian constraint generated with  the new elliptic
solver. We compare  with results for a certain  analytic approximation for the
initial data.  In the case of  three black holes, the  dynamics resulting from
approximate data is  different from the dynamics produced  by evolutions which
satisfy the  Hamiltonian constraint numerically. As  anticipated, the puncture
tracks  are sensitive to  small changes  in the  initial data.  Especially for
three and  more black holes  changing the initial  data, e.g.\ by  solving the
constraints  rather  than  using  an  analytical approximation,  can  lead  to
qualitatively  and quantitatively  very  different merger  sequences.  In  any
case,  we  confirmed  the  result of  \cite{CamLouZlo07f,LouZlo07a}  that,  as
expected,  the puncture  method lends  itself naturally  to the  simulation of
multiple black holes.

Multiple black hole evolutions  in numerical relativity demand highly accurate
methods for initial  data and evolution schemes.  Even  in Newtonian dynamics,
codes which calculate orbits  require sophisticated methods to maintain stable
and  accurate evolution over long  time scales.  Evolutions of  multiple black
holes could  be useful  as a  test case that  taxes the  accuracy of  codes in
comparatively short evolutions.

Simulations of  three, four, or  even more black  holes lead to  the following
question about more general merger situations: How can we determine the number
of  black holes involved  in a  merger from  the observation  of gravitational
waves? A first  analysis of this topic was given  previously using a Newtonian
approach \cite{TorHatAsa09}, with the perhaps surprising result that there are
certain degeneracies in the gravitational waves that prevent a trivial answer
to  this  question. In  the  future,  we plan  to  extend  our  research to  a
systematic study of the waveforms of multiple black hole configurations.

\acknowledgments

It is a pleasure to thank  Monique Baaske, David Hilditch, and Milton Ruiz for
valuable discussions and comments on  the manuscript.  This work was supported
in part by  DFG grant SFB/Transregio 7.  The computations  were carried out on
the HLRB2 system at the LRZ in Garching.

\appendix

\section{Convergence of high-order finite difference schemes for $C^n$ functions.}
\label{appendixA}

In certain cases,  the order of convergence of a  finite difference scheme can
be higher in the interior than at the boundary, without the lower order at the
boundary  spoiling   the  convergence  in   the  interior  (e.g.~\cite{Lev07},
Sec.\  2.12).   Here  we estimate  the  order  of  convergence of  a  standard
$p$-order finite difference scheme for an elliptic problem, where the solution
is  $C^\infty$  everywhere except  on  the origin  where  it  is $C^n$  (where
$n<p$). In order to simplify the notation, we will later restrict the examples
to the one  dimensional case. However, the extension  to the three dimensional
case is straightforward.

Let $\DLie$ be an elliptic operator, $\Omega \subset \R^3$ an open domain, and
$u:\Omega \rightarrow \R$ the solution of the problem
\begin{eqnarray} 
  \DLie u(\vec{x}) = \rho  \quad   &\mathrm{for}& \quad   
  \vec{x} \in \Omega,  \label{eqn:ExactElliptic}\\  
  \mathcal{B} u(\vec{x})  =  u_b(\vec{x}) \quad& \mathrm{for}& \quad 
  \vec{x} \in \partial \Omega, \label{eqn:ExactBound}
\end{eqnarray}
where $\mathcal{B}$ is a boundary  operator, $\rho:\Omega \rightarrow \R$ is a
source term, and $u \in C^\infty_{0}(\Omega) \cap C^n(0)$.
Let $\DLie^h$ be a finite difference representation of order $p$ of $\DLie$ in
a mesh  $\Omega^h \subset  \mathbb{N}^3$ with a  uniform grid size  $h$.  The
numerical solution $U^h:\Omega^h \rightarrow \R$ satisfies
\begin{eqnarray} 
  \DLie^h U^h(\vec{x}^h) = \rho^h(\vec{x}^h) \quad  &\mathrm{for}& \quad  
  \vec{x}^h \in \Omega^h,  \label{eqn:NumericElliptic}\\ 
  \mathcal{B}^h  U^h(\vec{x}^h)  = u^h_b(\vec{x}^h) \quad& \mathrm{for}& \quad
  \vec{x}^h \in \partial \Omega^h, \label{eqn:ExactBound2}
\end{eqnarray}
where $\mathcal{B}^h$ is a  discrete boundary operator and $\rho^h$ is
the restriction of $\rho$ on $\Omega^h$.

Given  a point  $x  \in \Omega$,  we  identify points  between $\Omega^h$  and
$\Omega$  by $x_i=x_0+i h$,  where $i  \in \{0,1,\dots,N  \}$. For  every grid
function  we  use  as   notation  $U_{i}:=U^h(x_i)$.   The  finite  difference
representation of $\DLie$ on the lattice location $x_i$ has for each direction
the form
\begin{equation}\label{eqn:finiteOperator}
\DLie^h U^h_{i} = \sum^{i+p}_{I=i-p} a_{I-i} U_{I},
\end{equation}
where the coefficients $a_{I-i}$ depend of the order of approximation and the
kind of stencil.   For example, the standard 2nd  order centered approximation
to the second derivative is defined by $a_{0}=-2/h^2$, $a_{\pm 1}=1/h^2$.

The truncation error is defined by
\begin{equation}\label{eqn:truncation}
\tau^h := \vert \DLie^h u^h - \rho^h \vert,
\end{equation}
where  $u^h$  is   the  restriction  of  $u$  to   the  grid  $\Omega^h$.  The
approximation has the  order of consistency $p > 0$ if  there is $h_0>0$ which
for all positive $h<h_0$ satisfies
\begin{equation}
\tau^h \leq C h^p,
\end{equation}
with  a  constant $C  >  0$  independent of  $h$.   The  standard approach  to
analyzing the error in a finite  difference approximation is to expand each of
the  function   values  of   $u^h$  in  a   Taylor  series  about   the  point
$(x_i)$. Taylor's theorem states that  for a function $u \in C^{n-1}([x_i,x])$
and $u \in C^{n}((x_i,x))$,
\begin{equation}
u(x)=\sum_{k=0}^{n-1} \frac{u^{(k)}_i}{k!}(x-x_i)^k +
\frac{u^{(n)}(\xi)}{n!}(x-x_i)^n,
\end{equation}
where $\xi \in [x_i,x]$ and $u^{(n)}$ denotes the $n$-th derivative.  For grid
functions the expansion formula is
\begin{equation}\label{eqn:TaylorGrid}
  u_j=\sum_{k=0}^{n-1} \frac{u^{(k)}_i}{k!}(j-i)^k h^k 
  + \frac{u^{(n)}(\xi)}{n!}(j-i)^n h^n.
\end{equation}
Using (\ref{eqn:finiteOperator}) and (\ref{eqn:TaylorGrid}), it is possible to
calculate
\begin{equation}
  \begin{split}
    \DLie^h u_i^h = &\sum^{i+p}_{I=i-p} \sum_{k=0}^{n-1} a_{I-i}
    \frac{u^{(k)}_i}{k!}(I-i)^k h^k \\ 
    & + \sum^{i+p}_{I=i-p} \sum_{k=0}^{n-1} a_{I-i} \frac{u^{(n)}(\xi)}{n!}(j-i)^n h^n.
  \end{split}
\end{equation}
If  $n \geq  p$ and  the  operator $\DLie$  contains a  linear combination  of
derivatives up to  order $n-1$, then it is possible  to select the coefficients
$a_{i}$ to cancel the remaining factors. We obtain
\begin{equation}
  \begin{split}
    \DLie^h u_i^h =&\DLie u_i^h +\sum^{i+p}_{I=i-p}\sum_{k=p}^{n-1} a_{I-i}   
    \frac{u^{(k)}_i}{k!}(I-i)^k h^k \\   
    &+\sum^{i+p}_{I=i-p}\sum_{k=0}^{n-1}a_{I-i}\frac{u^{(n)}(\xi)}{n!}(j-i)^n h^n,
  \end{split}
\end{equation}
where now the second summand starts at $k=p$. If $\vert u^{(n)}(\xi) \vert$ is
bounded,  the   dominant  term  is   of  order  $h^p$.  A   substitution  with
(\ref{eqn:truncation}) leads to
\begin{equation}\label{eqn:truncation2}
\tau^h \leq  \big \vert \sum^{i+p}_{I=i-p} a_{I-i} 
\frac{u^{(p)}_i}{p!}(I-i)^p \big \vert h^p,
\end{equation}
where the factor is bounded and independent  of $h$. If we use the same scheme
close to the origin,  where $n<p$, we are not able to  cancel terms lower than
$h^n$:
\begin{equation}
  \begin{split}
    \DLie^h  u_i^h =&  \DLie u_i^h  +  \sum^{i+p}_{I=i-p} \sum_{k=0}^{n-1}
    a_{I-i} \frac{u^{(n)}(\xi)}{n!}(j-i)^n h^n.
  \end{split}
\end{equation}
The truncation error in this case is of order $n<p$,
\begin{equation}\label{eqn:truncation3}
  \tau^h \leq \big \vert \sum^{i+p}_{I=i-p} a_{I-i}
  \frac{u^{(n)}(\xi)}{n!}(I-i)^n \big \vert h^n.
\end{equation}
For example, for the operator
\begin{equation}
  \DLie = \frac{\partial^2}{\partial x^2},
\end{equation}
the 4th order centered approximation to the second derivative is
\begin{equation}\label{eqn:secondDer}
  \DLie^h u^h_i = -\frac{u_{i-2}-16 u_{i-1}+30u_{i}-16u_{i+1}+u_{i+2}}{12 h^2}.
\end{equation}
If  $u \in  C^\infty_{0}(\R) \cap  C^2(0)$  and $0  \in [x_{i+1},x_{i+2}]$,  a
substitution  of the  Taylor series  of $x_{i  \pm 2}$  and $x_{i  \pm  1}$ in
equation (\ref{eqn:secondDer}) results in
\begin{equation}
  \begin{split}
    \DLie^h u^h_i =& \frac{\partial^2 u^h_i}{\partial x^2} + \frac{1}{9} h
    \left( \frac{\partial^3 u^h_i}{\partial x^3}
    -\frac{\partial^3 u^h(\xi)}{\partial x^3} \right) +\\
    &\frac{1}{18}h^2\frac{\partial^4 u^h(\xi)}{\partial x^4}+\mathcal{O}(h^3),
  \end{split}
\end{equation}
where  we  expand  the  term  $x_{i+2}$ only  up  to  $\mathcal{O}(h^2)$.  The
truncation  error  is of  order $\mathcal{O}(h)$ .  


\bibliography{refs,refs_extra}

\begin{thebibliography}{77}
\expandafter\ifx\csname natexlab\endcsname\relax\def\natexlab#1{#1}\fi
\expandafter\ifx\csname bibnamefont\endcsname\relax
  \def\bibnamefont#1{#1}\fi
\expandafter\ifx\csname bibfnamefont\endcsname\relax
  \def\bibfnamefont#1{#1}\fi
\expandafter\ifx\csname citenamefont\endcsname\relax
  \def\citenamefont#1{#1}\fi
\expandafter\ifx\csname url\endcsname\relax
  \def\url#1{\texttt{#1}}\fi
\expandafter\ifx\csname urlprefix\endcsname\relax\def\urlprefix{URL }\fi
\providecommand{\bibinfo}[2]{#2}
\providecommand{\eprint}[2][]{\url{#2}}

\bibitem[{\citenamefont{Valtonen and Karttunen}(2006)}]{ValHan06}
\bibinfo{author}{\bibfnamefont{M.~J.} \bibnamefont{Valtonen}} \bibnamefont{and}
  \bibinfo{author}{\bibfnamefont{H.}~\bibnamefont{Karttunen}},
  \emph{\bibinfo{title}{The three-body problem}} (\bibinfo{publisher}{Cambridge
  University Press}, \bibinfo{address}{New York}, \bibinfo{year}{2006}), ISBN
  \bibinfo{isbn}{0-521-85224-2 (hardcover)}.

\bibitem[{\citenamefont{{Laskar}}(1989)}]{Las89}
\bibinfo{author}{\bibfnamefont{J.}~\bibnamefont{{Laskar}}},
  \bibinfo{journal}{Nature} \textbf{\bibinfo{volume}{338}},
  \bibinfo{pages}{237} (\bibinfo{year}{1989}).

\bibitem[{\citenamefont{{Laskar}}(1994)}]{Las94}
\bibinfo{author}{\bibfnamefont{J.}~\bibnamefont{{Laskar}}},
  \bibinfo{journal}{Astron. Astrophys.} \textbf{\bibinfo{volume}{287}},
  \bibinfo{pages}{L9} (\bibinfo{year}{1994}).

\bibitem[{\citenamefont{Hayes}(2007)}]{Hay07}
\bibinfo{author}{\bibfnamefont{W.~B.} \bibnamefont{Hayes}},
  \bibinfo{journal}{Nat. Phys.} \textbf{\bibinfo{volume}{3}},
  \bibinfo{pages}{1745} (\bibinfo{year}{2007}),
  \urlprefix\url{http://dx.doi.org/10.1038/nphys728}.

\bibitem[{\citenamefont{Gultekin et~al.}(2003)\citenamefont{Gultekin, Miller,
  and Hamilton}}]{Gul03}
\bibinfo{author}{\bibfnamefont{K.}~\bibnamefont{Gultekin}},
  \bibinfo{author}{\bibfnamefont{M.~C.} \bibnamefont{Miller}},
  \bibnamefont{and} \bibinfo{author}{\bibfnamefont{D.~P.}
  \bibnamefont{Hamilton}}, \bibinfo{journal}{AIP Conf. Proc.}
  \textbf{\bibinfo{volume}{686}}, \bibinfo{pages}{135} (\bibinfo{year}{2003}),
  \eprint{astro-ph/0306204}.

\bibitem[{\citenamefont{{Coleman Miller}}(2003)}]{ColMil03}
\bibinfo{author}{\bibfnamefont{M.}~\bibnamefont{{Coleman Miller}}},
  \bibinfo{journal}{AIP Conf. Proc.} \textbf{\bibinfo{volume}{686}},
  \bibinfo{pages}{125} (\bibinfo{year}{2003}), \eprint{astro-ph/0306173}.

\bibitem[{\citenamefont{Gultekin et~al.}(2004)\citenamefont{Gultekin, Miller,
  and Hamilton}}]{GulMilHam04}
\bibinfo{author}{\bibfnamefont{K.}~\bibnamefont{Gultekin}},
  \bibinfo{author}{\bibfnamefont{M.~C.} \bibnamefont{Miller}},
  \bibnamefont{and} \bibinfo{author}{\bibfnamefont{D.~P.}
  \bibnamefont{Hamilton}}, \bibinfo{journal}{ApJ}
  \textbf{\bibinfo{volume}{616}}, \bibinfo{pages}{221} (\bibinfo{year}{2004}).

\bibitem[{\citenamefont{Valtonen and Mikkola}(1991)}]{ValMik91}
\bibinfo{author}{\bibfnamefont{M.}~\bibnamefont{Valtonen}} \bibnamefont{and}
  \bibinfo{author}{\bibfnamefont{S.}~\bibnamefont{Mikkola}},
  \bibinfo{journal}{Annual Review of Astronomy and Astrophysics}
  \textbf{\bibinfo{volume}{29}}, \bibinfo{pages}{9} (\bibinfo{year}{1991}).

\bibitem[{\citenamefont{{Portegies Zwart} and McMillan}(2000)}]{ZwaMcM00}
\bibinfo{author}{\bibfnamefont{S.~F.} \bibnamefont{{Portegies Zwart}}}
  \bibnamefont{and} \bibinfo{author}{\bibfnamefont{S.~L.~W.}
  \bibnamefont{McMillan}}, \bibinfo{journal}{Astrophys. J. Lett.}
  \textbf{\bibinfo{volume}{528}}, \bibinfo{pages}{L17} (\bibinfo{year}{2000}).

\bibitem[{\citenamefont{Gualandris et~al.}(2005)\citenamefont{Gualandris,
  Zwart, and Sipior}}]{GuaZwaSip05}
\bibinfo{author}{\bibfnamefont{A.}~\bibnamefont{Gualandris}},
  \bibinfo{author}{\bibfnamefont{S.~P.} \bibnamefont{Zwart}}, \bibnamefont{and}
  \bibinfo{author}{\bibfnamefont{M.}~\bibnamefont{Sipior}},
  \bibinfo{journal}{Mon. Not. R. Astron. Soc.} \textbf{\bibinfo{volume}{363}},
  \bibinfo{pages}{223} (\bibinfo{year}{2005}).

\bibitem[{\citenamefont{Preto et~al.}(2009)\citenamefont{Preto, Berentzen,
  Berczik, Merritt, and Spurzem}}]{PreBerBer08}
\bibinfo{author}{\bibfnamefont{M.}~\bibnamefont{Preto}},
  \bibinfo{author}{\bibfnamefont{I.}~\bibnamefont{Berentzen}},
  \bibinfo{author}{\bibfnamefont{P.}~\bibnamefont{Berczik}},
  \bibinfo{author}{\bibfnamefont{D.}~\bibnamefont{Merritt}}, \bibnamefont{and}
  \bibinfo{author}{\bibfnamefont{R.}~\bibnamefont{Spurzem}},
  \bibinfo{journal}{J. Phys. Conf. Ser.} \textbf{\bibinfo{volume}{154}},
  \bibinfo{pages}{012049} (\bibinfo{year}{2009}), \eprint{0811.3501}.

\bibitem[{\citenamefont{Springel et~al.}(2005)\citenamefont{Springel, White,
  Jenkins, Frenk, Yoshida, Gao, Navarro, Thacker, Croton, Helly
  et~al.}}]{VolEtal05}
\bibinfo{author}{\bibfnamefont{V.}~\bibnamefont{Springel}},
  \bibinfo{author}{\bibfnamefont{S.~D.~M.} \bibnamefont{White}},
  \bibinfo{author}{\bibfnamefont{A.}~\bibnamefont{Jenkins}},
  \bibinfo{author}{\bibfnamefont{C.~S.} \bibnamefont{Frenk}},
  \bibinfo{author}{\bibfnamefont{N.}~\bibnamefont{Yoshida}},
  \bibinfo{author}{\bibfnamefont{L.}~\bibnamefont{Gao}},
  \bibinfo{author}{\bibfnamefont{J.}~\bibnamefont{Navarro}},
  \bibinfo{author}{\bibfnamefont{R.}~\bibnamefont{Thacker}},
  \bibinfo{author}{\bibfnamefont{D.}~\bibnamefont{Croton}},
  \bibinfo{author}{\bibfnamefont{J.}~\bibnamefont{Helly}},
  \bibnamefont{et~al.}, \bibinfo{journal}{Nature}
  \textbf{\bibinfo{volume}{435}}, \bibinfo{pages}{629} (\bibinfo{year}{2005}),
  \bibinfo{note}{astro-ph/0504097}.

\bibitem[{\citenamefont{Bertschinger}(1998)}]{Ber98}
\bibinfo{author}{\bibfnamefont{E.}~\bibnamefont{Bertschinger}},
  \bibinfo{journal}{Annu. Rev. Astron. Astrophys.}
  \textbf{\bibinfo{volume}{36}}, \bibinfo{pages}{599} (\bibinfo{year}{1998}).

\bibitem[{\citenamefont{Hatton et~al.}(2003)\citenamefont{Hatton, Devriendt,
  Ninin, Bouchet, Guiderdoni, and Vibert}}]{Hat03}
\bibinfo{author}{\bibfnamefont{S.}~\bibnamefont{Hatton}},
  \bibinfo{author}{\bibfnamefont{J.~E.~G.} \bibnamefont{Devriendt}},
  \bibinfo{author}{\bibfnamefont{S.}~\bibnamefont{Ninin}},
  \bibinfo{author}{\bibfnamefont{F.~R.} \bibnamefont{Bouchet}},
  \bibinfo{author}{\bibfnamefont{B.}~\bibnamefont{Guiderdoni}},
  \bibnamefont{and} \bibinfo{author}{\bibfnamefont{D.}~\bibnamefont{Vibert}},
  \bibinfo{journal}{Mon. Not. Roy. Astron. Soc.}
  \textbf{\bibinfo{volume}{343}}, \bibinfo{pages}{75} (\bibinfo{year}{2003}),
  \eprint{astro-ph/0309186}.

\bibitem[{\citenamefont{Fregeau et~al.}(2004)\citenamefont{Fregeau, Cheung,
  Zwart, and Rasio}}]{FreCheZwa04}
\bibinfo{author}{\bibfnamefont{J.~M.} \bibnamefont{Fregeau}},
  \bibinfo{author}{\bibfnamefont{P.}~\bibnamefont{Cheung}},
  \bibinfo{author}{\bibfnamefont{S.~F.~P.} \bibnamefont{Zwart}},
  \bibnamefont{and} \bibinfo{author}{\bibfnamefont{F.~A.} \bibnamefont{Rasio}},
  \bibinfo{journal}{Mon. Not. R Astron. Soc.} \textbf{\bibinfo{volume}{352}},
  \bibinfo{pages}{1} (\bibinfo{year}{2004}).

\bibitem[{\citenamefont{Goldstein et~al.}(2001)\citenamefont{Goldstein, Poole,
  and Safko}}]{GolPooSaf01}
\bibinfo{author}{\bibfnamefont{H.}~\bibnamefont{Goldstein}},
  \bibinfo{author}{\bibfnamefont{C.~P.} \bibnamefont{Poole}}, \bibnamefont{and}
  \bibinfo{author}{\bibfnamefont{J.}~\bibnamefont{Safko}},
  \emph{\bibinfo{title}{Classical Mechanics}} (\bibinfo{publisher}{Addison
  Wesley}, \bibinfo{year}{2001}), ISBN \bibinfo{isbn}{0-201-65702-3}.

\bibitem[{\citenamefont{Sundman}(1907)}]{Sun1907}
\bibinfo{author}{\bibfnamefont{K.}~\bibnamefont{Sundman}},
  \bibinfo{journal}{Acta Soc. Sci. Fennicae}  (\bibinfo{year}{1907}).

\bibitem[{\citenamefont{Barrow-Green}(1996)}]{Bar96}
\bibinfo{author}{\bibfnamefont{J.}~\bibnamefont{Barrow-Green}},
  \emph{\bibinfo{title}{Poincare and the Three Body Problem}}
  (\bibinfo{publisher}{American Mathematical Society}, \bibinfo{year}{1996}),
  ISBN \bibinfo{isbn}{0821803670}.

\bibitem[{\citenamefont{Jaranowski and Sch{\"a}fer}(1997)}]{JarSch97}
\bibinfo{author}{\bibfnamefont{P.}~\bibnamefont{Jaranowski}} \bibnamefont{and}
  \bibinfo{author}{\bibfnamefont{G.}~\bibnamefont{Sch{\"a}fer}},
  \bibinfo{journal}{Phys. Rev. D} \textbf{\bibinfo{volume}{55}},
  \bibinfo{pages}{4712} (\bibinfo{year}{1997}).

\bibitem[{\citenamefont{Blanchet}(2002)}]{Bla02}
\bibinfo{author}{\bibfnamefont{L.}~\bibnamefont{Blanchet}},
  \bibinfo{journal}{Living Rev. Relativity} \textbf{\bibinfo{volume}{5}},
  \bibinfo{pages}{3} (\bibinfo{year}{2002}), \eprint{gr-qc/0202016},
  \urlprefix\url{http://www.livingreviews.org/lrr-2002-3}.

\bibitem[{\citenamefont{{Toshifumi Futamase}}(2007)}]{FutIto07}
\bibinfo{author}{\bibfnamefont{Y.~I.} \bibnamefont{{Toshifumi Futamase}}},
  \bibinfo{journal}{Living Reviews in Relativity} \textbf{\bibinfo{volume}{10}}
  (\bibinfo{year}{2007}),
  \urlprefix\url{http://www.livingreviews.org/lrr-2007-2}.

\bibitem[{\citenamefont{K{\"o}nigsd{\"o}rffer
  et~al.}(2003)\citenamefont{K{\"o}nigsd{\"o}rffer, Faye, and
  Sch{\"a}fer}}]{KonFaySch03}
\bibinfo{author}{\bibfnamefont{C.}~\bibnamefont{K{\"o}nigsd{\"o}rffer}},
  \bibinfo{author}{\bibfnamefont{G.}~\bibnamefont{Faye}}, \bibnamefont{and}
  \bibinfo{author}{\bibfnamefont{G.}~\bibnamefont{Sch{\"a}fer}},
  \bibinfo{journal}{Phys. Rev. D} \textbf{\bibinfo{volume}{68}},
  \bibinfo{pages}{044004} (\bibinfo{year}{2003}).

\bibitem[{\citenamefont{Chu}(2009)}]{Chu09}
\bibinfo{author}{\bibfnamefont{Y.-Z.} \bibnamefont{Chu}},
  \bibinfo{journal}{Phys. Rev. D} \textbf{\bibinfo{volume}{79}},
  \bibinfo{pages}{044031} (\bibinfo{year}{2009}).

\bibitem[{\citenamefont{Sch{\"a}fer}(1987)}]{Sch87}
\bibinfo{author}{\bibfnamefont{G.}~\bibnamefont{Sch{\"a}fer}},
  \bibinfo{journal}{Physics Letters A} \textbf{\bibinfo{volume}{123}},
  \bibinfo{pages}{336} (\bibinfo{year}{1987}).

\bibitem[{\citenamefont{Lousto and Nakano}(2008)}]{LouHir08}
\bibinfo{author}{\bibfnamefont{C.~O.} \bibnamefont{Lousto}} \bibnamefont{and}
  \bibinfo{author}{\bibfnamefont{H.}~\bibnamefont{Nakano}},
  \bibinfo{journal}{Classical and Quantum Gravity}
  \textbf{\bibinfo{volume}{25}}, \bibinfo{pages}{195019}
  (\bibinfo{year}{2008}).

\bibitem[{\citenamefont{Moore}(1993)}]{Moo93}
\bibinfo{author}{\bibfnamefont{C.}~\bibnamefont{Moore}},
  \bibinfo{journal}{Phys. Rev. Lett.} \textbf{\bibinfo{volume}{70}},
  \bibinfo{pages}{3675} (\bibinfo{year}{1993}).

\bibitem[{\citenamefont{Imai et~al.}(2007)\citenamefont{Imai, Chiba, and
  Asada}}]{TatTakHid07}
\bibinfo{author}{\bibfnamefont{T.}~\bibnamefont{Imai}},
  \bibinfo{author}{\bibfnamefont{T.}~\bibnamefont{Chiba}}, \bibnamefont{and}
  \bibinfo{author}{\bibfnamefont{H.}~\bibnamefont{Asada}},
  \bibinfo{journal}{Phys. Rev. Lett.} \textbf{\bibinfo{volume}{98}},
  \bibinfo{pages}{201102} (\bibinfo{year}{2007}).

\bibitem[{\citenamefont{G{\"u}ltekin et~al.}(2003)\citenamefont{G{\"u}ltekin,
  Miller, and Hamilton}}]{GulMilHam05}
\bibinfo{author}{\bibfnamefont{K.}~\bibnamefont{G{\"u}ltekin}},
  \bibinfo{author}{\bibfnamefont{M.~C.} \bibnamefont{Miller}},
  \bibnamefont{and} \bibinfo{author}{\bibfnamefont{D.~P.}
  \bibnamefont{Hamilton}} (\bibinfo{publisher}{AIP}, \bibinfo{year}{2003}),
  vol. \bibinfo{volume}{686}, pp. \bibinfo{pages}{135--140}.

\bibitem[{\citenamefont{Campanelli
  et~al.}(2006{\natexlab{a}})\citenamefont{Campanelli, Dettwyler, Hannam, and
  Lousto}}]{CamDetHan05}
\bibinfo{author}{\bibfnamefont{M.}~\bibnamefont{Campanelli}},
  \bibinfo{author}{\bibfnamefont{M.}~\bibnamefont{Dettwyler}},
  \bibinfo{author}{\bibfnamefont{M.}~\bibnamefont{Hannam}}, \bibnamefont{and}
  \bibinfo{author}{\bibfnamefont{C.~O.} \bibnamefont{Lousto}},
  \bibinfo{journal}{Phys. Rev. D} \textbf{\bibinfo{volume}{74}},
  \bibinfo{pages}{087503} (\bibinfo{year}{2006}{\natexlab{a}}),
  \eprint{astro-ph/0509814}.

\bibitem[{\citenamefont{Campanelli et~al.}(2008)\citenamefont{Campanelli,
  Lousto, and Zlochower}}]{CamLouZlo07f}
\bibinfo{author}{\bibfnamefont{M.}~\bibnamefont{Campanelli}},
  \bibinfo{author}{\bibfnamefont{C.~O.} \bibnamefont{Lousto}},
  \bibnamefont{and}
  \bibinfo{author}{\bibfnamefont{Y.}~\bibnamefont{Zlochower}},
  \bibinfo{journal}{Phys. Rev.} \textbf{\bibinfo{volume}{D77}},
  \bibinfo{pages}{101501} (\bibinfo{year}{2008}), \eprint{0710.0879}.

\bibitem[{\citenamefont{Lousto and Zlochower}(2008)}]{LouZlo07a}
\bibinfo{author}{\bibfnamefont{C.~O.} \bibnamefont{Lousto}} \bibnamefont{and}
  \bibinfo{author}{\bibfnamefont{Y.}~\bibnamefont{Zlochower}},
  \bibinfo{journal}{Phys. Rev.} \textbf{\bibinfo{volume}{D77}},
  \bibinfo{pages}{024034} (\bibinfo{year}{2008}), \eprint{arXiv:0711.1165
  [gr-qc]}.

\bibitem[{\citenamefont{Br{\"u}gmann}(1997)}]{Bru97a}
\bibinfo{author}{\bibfnamefont{B.}~\bibnamefont{Br{\"u}gmann}},
  \emph{\bibinfo{title}{Evolution of 30 black holes spelling {AEI}}}
  (\bibinfo{year}{1997}), \bibinfo{note}{{T}alk for Fachbeirat, Albert Einstein
  Institute. Unpublished.}

\bibitem[{\citenamefont{Diener}(2003)}]{Die03}
\bibinfo{author}{\bibfnamefont{P.}~\bibnamefont{Diener}},
  \bibinfo{journal}{Class. Quantum Grav.} \textbf{\bibinfo{volume}{20}},
  \bibinfo{pages}{4901} (\bibinfo{year}{2003}), \eprint{gr-qc/0305039}.

\bibitem[{Bru()}]{Bru97atext}
\bibinfo{note}{The first proof of principle simulation showing that puncture
  evolutions generalize to three or more black holes with minimal changes to a
  binary code was performed in 1997~\cite{Bru97a}. Since this was an
  unpublished report, we summarize one of these simulations here. 30 black
  holes were arranged in a planar configuration using Brill-Lindquist data.
  Evolutions were performed using the fixed puncture method with the ADM
  formulation, maximal slicing, and vanishing shift, using an early version of
  the~\textsc{BAM} code~\cite{Bru96,Bru97}. Shown at \cite{emc2} is the lapse
  at $t=0.5M$, which was initialized to one and collapsed quickly towards zero
  near the punctures, thereby marking the location of the black holes. These
  simulations were not stable on orbital time scales, so neither the full
  merger nor waveforms were computed. About at the same time, there were also
  experiments with three black holes using the Cactus code, for which we are
  only aware of reference~\cite{Die03}.}

\bibitem[{\citenamefont{Br{\"u}gmann}(1996)}]{Bru96}
\bibinfo{author}{\bibfnamefont{B.}~\bibnamefont{Br{\"u}gmann}},
  \bibinfo{journal}{Phys. Rev. D} \textbf{\bibinfo{volume}{54}},
  \bibinfo{pages}{7361} (\bibinfo{year}{1996}), \eprint{gr-qc/9608050}.

\bibitem[{\citenamefont{Br{\"u}gmann}(1999)}]{Bru97}
\bibinfo{author}{\bibfnamefont{B.}~\bibnamefont{Br{\"u}gmann}},
  \bibinfo{journal}{Int. J. Mod. Phys.} \textbf{\bibinfo{volume}{8}},
  \bibinfo{pages}{85} (\bibinfo{year}{1999}), \eprint{gr-qc/9708035}.

\bibitem[{emc()}]{emc2}
\urlprefix\url{http://www.tpi.uni-jena.de/gravity/Showcase/}.

\bibitem[{\citenamefont{Beig and O'Murchadha}(1994)}]{BeiOMu94}
\bibinfo{author}{\bibfnamefont{R.}~\bibnamefont{Beig}} \bibnamefont{and}
  \bibinfo{author}{\bibfnamefont{N.}~\bibnamefont{O'Murchadha}},
  \bibinfo{journal}{Class. Quantum Grav.} \textbf{\bibinfo{volume}{11}},
  \bibinfo{pages}{419} (\bibinfo{year}{1994}).

\bibitem[{\citenamefont{Beig and O'Murchadha}(1996)}]{BeiOMu96}
\bibinfo{author}{\bibfnamefont{R.}~\bibnamefont{Beig}} \bibnamefont{and}
  \bibinfo{author}{\bibfnamefont{N.}~\bibnamefont{O'Murchadha}},
  \bibinfo{journal}{Class. Quantum Grav.} \textbf{\bibinfo{volume}{13}},
  \bibinfo{pages}{739} (\bibinfo{year}{1996}).

\bibitem[{\citenamefont{Brandt and Br{\"u}gmann}(1997)}]{BraBru97}
\bibinfo{author}{\bibfnamefont{S.}~\bibnamefont{Brandt}} \bibnamefont{and}
  \bibinfo{author}{\bibfnamefont{B.}~\bibnamefont{Br{\"u}gmann}},
  \bibinfo{journal}{Phys. Rev. Lett.} \textbf{\bibinfo{volume}{78}},
  \bibinfo{pages}{3606} (\bibinfo{year}{1997}), \eprint{gr-qc/9703066}.

\bibitem[{\citenamefont{Cao et~al.}(2008)\citenamefont{Cao, Yo, and
  Yu}}]{CaoYoYu08}
\bibinfo{author}{\bibfnamefont{Z.}~\bibnamefont{Cao}},
  \bibinfo{author}{\bibfnamefont{H.-J.} \bibnamefont{Yo}}, \bibnamefont{and}
  \bibinfo{author}{\bibfnamefont{J.-P.} \bibnamefont{Yu}},
  \bibinfo{journal}{Phys. Rev. D} \textbf{\bibinfo{volume}{78}},
  \bibinfo{pages}{124011} (\bibinfo{year}{2008}).

\bibitem[{\citenamefont{Alcubierre}(2008)}]{Alc08}
\bibinfo{author}{\bibfnamefont{M.}~\bibnamefont{Alcubierre}},
  \emph{\bibinfo{title}{Introduction to 3+1 Numerical Relativity}},
  International Series of Monographs on Physics (\bibinfo{publisher}{Oxford
  University Press, USA}, \bibinfo{year}{2008}).

\bibitem[{\citenamefont{Cook}(2000)}]{Coo00}
\bibinfo{author}{\bibfnamefont{G.~B.} \bibnamefont{Cook}},
  \bibinfo{journal}{Living Rev. Relativity} \textbf{\bibinfo{volume}{3}},
  \bibinfo{pages}{5} (\bibinfo{year}{2000}),
  \urlprefix\url{http://www.livingreviews.org/lrr-2000-5}.

\bibitem[{\citenamefont{Gourgoulhon}(2007)}]{Gou07b}
\bibinfo{author}{\bibfnamefont{E.}~\bibnamefont{Gourgoulhon}},
  \bibinfo{journal}{Journal of Physics: Conference Series}
  \textbf{\bibinfo{volume}{91}}, \bibinfo{pages}{012001}
  (\bibinfo{year}{2007}).

\bibitem[{\citenamefont{Brill and Lindquist}(1963)}]{BriLin63}
\bibinfo{author}{\bibfnamefont{D.~R.} \bibnamefont{Brill}} \bibnamefont{and}
  \bibinfo{author}{\bibfnamefont{R.~W.} \bibnamefont{Lindquist}},
  \bibinfo{journal}{Phys. Rev.} \textbf{\bibinfo{volume}{131}},
  \bibinfo{pages}{471} (\bibinfo{year}{1963}).

\bibitem[{\citenamefont{Bowen and York}(1980)}]{BowYor80}
\bibinfo{author}{\bibfnamefont{J.~M.} \bibnamefont{Bowen}} \bibnamefont{and}
  \bibinfo{author}{\bibfnamefont{J.~W.} \bibnamefont{York},
  \bibfnamefont{Jr.}}, \bibinfo{journal}{Phys. Rev. D}
  \textbf{\bibinfo{volume}{21}}, \bibinfo{pages}{2047} (\bibinfo{year}{1980}).

\bibitem[{\citenamefont{Brandt}(1977)}]{Bra77}
\bibinfo{author}{\bibfnamefont{A.}~\bibnamefont{Brandt}},
  \bibinfo{journal}{Math. Comp.} \textbf{\bibinfo{volume}{31}},
  \bibinfo{pages}{333} (\bibinfo{year}{1977}).

\bibitem[{\citenamefont{Bai and Brandt}(1987)}]{BaiBra87}
\bibinfo{author}{\bibfnamefont{D.}~\bibnamefont{Bai}} \bibnamefont{and}
  \bibinfo{author}{\bibfnamefont{A.}~\bibnamefont{Brandt}},
  \bibinfo{journal}{SIAM J. Sci. Stat. Comput.} \textbf{\bibinfo{volume}{8}},
  \bibinfo{pages}{109} (\bibinfo{year}{1987}).

\bibitem[{\citenamefont{Brandt and Lanza}(1988)}]{BraLan88}
\bibinfo{author}{\bibfnamefont{A.}~\bibnamefont{Brandt}} \bibnamefont{and}
  \bibinfo{author}{\bibfnamefont{A.}~\bibnamefont{Lanza}},
  \bibinfo{journal}{Class. Quantum Grav.} \textbf{\bibinfo{volume}{5}},
  \bibinfo{pages}{713} (\bibinfo{year}{1988}).

\bibitem[{\citenamefont{Hawley and Matzner}(2004)}]{HawMat03}
\bibinfo{author}{\bibfnamefont{S.~H.} \bibnamefont{Hawley}} \bibnamefont{and}
  \bibinfo{author}{\bibfnamefont{R.~A.} \bibnamefont{Matzner}},
  \bibinfo{journal}{Class. Quantum Grav.} \textbf{\bibinfo{volume}{21}},
  \bibinfo{pages}{805} (\bibinfo{year}{2004}), \eprint{gr-qc/0306122}.

\bibitem[{\citenamefont{Choptuik and Unruh}(1986)}]{ChoUnr86b}
\bibinfo{author}{\bibfnamefont{M.~W.} \bibnamefont{Choptuik}} \bibnamefont{and}
  \bibinfo{author}{\bibfnamefont{W.~G.} \bibnamefont{Unruh}},
  \bibinfo{journal}{Gen. Rel. Grav.} \textbf{\bibinfo{volume}{18}},
  \bibinfo{pages}{813} (\bibinfo{year}{1986}).

\bibitem[{\citenamefont{Choptuik}(2006)}]{Cho06a}
\bibinfo{author}{\bibfnamefont{M.~W.} \bibnamefont{Choptuik}}
  (\bibinfo{year}{2006}), \bibinfo{note}{course given at VII Mexican School on
  Gravitation and Mathematical Physics, November 26 2006},
  \urlprefix\url{www.smf.mx/\~dgfm-smf/EscuelaVII/courses.html}.

\bibitem[{\citenamefont{Linz}(2001)}]{Lin01}
\bibinfo{author}{\bibfnamefont{P.}~\bibnamefont{Linz}},
  \emph{\bibinfo{title}{Theoretical Numerical Analysis}}
  (\bibinfo{publisher}{Dover Publications}, \bibinfo{year}{2001}).

\bibitem[{\citenamefont{Ansorg et~al.}(2004)\citenamefont{Ansorg, Br{\"u}gmann,
  and Tichy}}]{AnsBruTic04}
\bibinfo{author}{\bibfnamefont{M.}~\bibnamefont{Ansorg}},
  \bibinfo{author}{\bibfnamefont{B.}~\bibnamefont{Br{\"u}gmann}},
  \bibnamefont{and} \bibinfo{author}{\bibfnamefont{W.}~\bibnamefont{Tichy}},
  \bibinfo{journal}{Phys. Rev. D} \textbf{\bibinfo{volume}{70}},
  \bibinfo{pages}{064011} (\bibinfo{year}{2004}), \eprint{gr-qc/0404056}.

\bibitem[{\citenamefont{Boyd}(2001)}]{boy01}
\bibinfo{author}{\bibfnamefont{J.~P.} \bibnamefont{Boyd}},
  \emph{\bibinfo{title}{Chebyshev and Fourier Spectral Methods (Second Edition,
  Revised)}} (\bibinfo{publisher}{Dover Publications}, \bibinfo{address}{New
  York}, \bibinfo{year}{2001}), ISBN \bibinfo{isbn}{0-486-41183-4}.

\bibitem[{\citenamefont{Laguna}(2004)}]{Lag03a}
\bibinfo{author}{\bibfnamefont{P.}~\bibnamefont{Laguna}},
  \bibinfo{journal}{Phys. Rev. D} \textbf{\bibinfo{volume}{69}},
  \bibinfo{pages}{104020} (\bibinfo{year}{2004}), \eprint{gr-qc/0310073}.

\bibitem[{\citenamefont{Dennison et~al.}(2006)\citenamefont{Dennison,
  Baumgarte, and Pfeiffer}}]{DenBauPfe06}
\bibinfo{author}{\bibfnamefont{K.~A.} \bibnamefont{Dennison}},
  \bibinfo{author}{\bibfnamefont{T.~W.} \bibnamefont{Baumgarte}},
  \bibnamefont{and} \bibinfo{author}{\bibfnamefont{H.~P.}
  \bibnamefont{Pfeiffer}}, \bibinfo{journal}{Phys. Rev.}
  \textbf{\bibinfo{volume}{D74}}, \bibinfo{pages}{064016}
  (\bibinfo{year}{2006}), \eprint{gr-qc/0606037}.

\bibitem[{\citenamefont{Gleiser et~al.}(2002)\citenamefont{Gleiser, Khanna, and
  Pullin}}]{GleKhaPul99}
\bibinfo{author}{\bibfnamefont{R.~J.} \bibnamefont{Gleiser}},
  \bibinfo{author}{\bibfnamefont{G.}~\bibnamefont{Khanna}}, \bibnamefont{and}
  \bibinfo{author}{\bibfnamefont{J.}~\bibnamefont{Pullin}},
  \bibinfo{journal}{Phys. Rev. D} \textbf{\bibinfo{volume}{66}},
  \bibinfo{pages}{024035} (\bibinfo{year}{2002}), \eprint{gr-qc/9905067}.

\bibitem[{\citenamefont{Gleiser et~al.}(1998)\citenamefont{Gleiser, Nicasio,
  Price, and Pullin}}]{GleNicPri97}
\bibinfo{author}{\bibfnamefont{R.~J.} \bibnamefont{Gleiser}},
  \bibinfo{author}{\bibfnamefont{C.~O.} \bibnamefont{Nicasio}},
  \bibinfo{author}{\bibfnamefont{R.~H.} \bibnamefont{Price}}, \bibnamefont{and}
  \bibinfo{author}{\bibfnamefont{J.}~\bibnamefont{Pullin}},
  \bibinfo{journal}{Phys. Rev. D} \textbf{\bibinfo{volume}{57}},
  \bibinfo{pages}{3401} (\bibinfo{year}{1998}), \eprint{gr-qc/9710096}.

\bibitem[{\citenamefont{Hahn and Lindquist}(1964)}]{HahLin64}
\bibinfo{author}{\bibfnamefont{S.~G.} \bibnamefont{Hahn}} \bibnamefont{and}
  \bibinfo{author}{\bibfnamefont{R.~W.} \bibnamefont{Lindquist}},
  \bibinfo{journal}{Ann. Phys.} \textbf{\bibinfo{volume}{29}},
  \bibinfo{pages}{304} (\bibinfo{year}{1964}).

\bibitem[{\citenamefont{Pretorius}(2005)}]{pre05}
\bibinfo{author}{\bibfnamefont{F.}~\bibnamefont{Pretorius}},
  \bibinfo{journal}{Phys. Rev. Lett.} \textbf{\bibinfo{volume}{95}},
  \bibinfo{pages}{121101} (\bibinfo{year}{2005}), \eprint{gr-qc/0507014}.

\bibitem[{\citenamefont{Campanelli
  et~al.}(2006{\natexlab{b}})\citenamefont{Campanelli, Lousto, Marronetti, and
  Zlochower}}]{CamLouMar05}
\bibinfo{author}{\bibfnamefont{M.}~\bibnamefont{Campanelli}},
  \bibinfo{author}{\bibfnamefont{C.~O.} \bibnamefont{Lousto}},
  \bibinfo{author}{\bibfnamefont{P.}~\bibnamefont{Marronetti}},
  \bibnamefont{and}
  \bibinfo{author}{\bibfnamefont{Y.}~\bibnamefont{Zlochower}},
  \bibinfo{journal}{Phys. Rev. Lett.} \textbf{\bibinfo{volume}{96}},
  \bibinfo{pages}{111101} (\bibinfo{year}{2006}{\natexlab{b}}),
  \eprint{gr-qc/0511048}.

\bibitem[{\citenamefont{Baker et~al.}(2006)\citenamefont{Baker, Centrella,
  Choi, Koppitz, and van Meter}}]{BakCenCho05b}
\bibinfo{author}{\bibfnamefont{J.~G.} \bibnamefont{Baker}},
  \bibinfo{author}{\bibfnamefont{J.}~\bibnamefont{Centrella}},
  \bibinfo{author}{\bibfnamefont{D.-I.} \bibnamefont{Choi}},
  \bibinfo{author}{\bibfnamefont{M.}~\bibnamefont{Koppitz}}, \bibnamefont{and}
  \bibinfo{author}{\bibfnamefont{J.}~\bibnamefont{van Meter}},
  \bibinfo{journal}{Phys. Rev. Lett.} \textbf{\bibinfo{volume}{96}},
  \bibinfo{pages}{111102} (\bibinfo{year}{2006}), \eprint{gr-qc/0511103}.

\bibitem[{\citenamefont{Br{\"u}gmann et~al.}(2004)\citenamefont{Br{\"u}gmann,
  Tichy, and Jansen}}]{BruTicJan03}
\bibinfo{author}{\bibfnamefont{B.}~\bibnamefont{Br{\"u}gmann}},
  \bibinfo{author}{\bibfnamefont{W.}~\bibnamefont{Tichy}}, \bibnamefont{and}
  \bibinfo{author}{\bibfnamefont{N.}~\bibnamefont{Jansen}},
  \bibinfo{journal}{Phys. Rev. Lett.} \textbf{\bibinfo{volume}{92}},
  \bibinfo{pages}{211101} (\bibinfo{year}{2004}), \eprint{gr-qc/0312112}.

\bibitem[{\citenamefont{Scheel et~al.}(2006)\citenamefont{Scheel, Pfeiffer,
  Lindblom, Kidder, Rinne, and Teukolsky}}]{SchPfeLin06}
\bibinfo{author}{\bibfnamefont{M.~A.} \bibnamefont{Scheel}},
  \bibinfo{author}{\bibfnamefont{H.~P.} \bibnamefont{Pfeiffer}},
  \bibinfo{author}{\bibfnamefont{L.}~\bibnamefont{Lindblom}},
  \bibinfo{author}{\bibfnamefont{L.~E.} \bibnamefont{Kidder}},
  \bibinfo{author}{\bibfnamefont{O.}~\bibnamefont{Rinne}}, \bibnamefont{and}
  \bibinfo{author}{\bibfnamefont{S.~A.} \bibnamefont{Teukolsky}},
  \bibinfo{journal}{Phys. Rev. D} \textbf{\bibinfo{volume}{74}},
  \bibinfo{pages}{104006} (\bibinfo{year}{2006}), \eprint{gr-qc/0607056}.

\bibitem[{\citenamefont{Shibata and Nakamura}(1995)}]{ShiNak95a}
\bibinfo{author}{\bibfnamefont{M.}~\bibnamefont{Shibata}} \bibnamefont{and}
  \bibinfo{author}{\bibfnamefont{T.}~\bibnamefont{Nakamura}},
  \bibinfo{journal}{Phys. Rev. D} \textbf{\bibinfo{volume}{52}},
  \bibinfo{pages}{5428} (\bibinfo{year}{1995}).

\bibitem[{\citenamefont{Baumgarte and Shapiro}(1998)}]{BauSha98b}
\bibinfo{author}{\bibfnamefont{T.~W.} \bibnamefont{Baumgarte}}
  \bibnamefont{and} \bibinfo{author}{\bibfnamefont{S.~L.}
  \bibnamefont{Shapiro}}, \bibinfo{journal}{Phys. Rev. D}
  \textbf{\bibinfo{volume}{59}}, \bibinfo{pages}{024007}
  (\bibinfo{year}{1998}), \eprint{gr-qc/9810065}.

\bibitem[{\citenamefont{Br{\"u}gmann et~al.}(2008)\citenamefont{Br{\"u}gmann,
  Gonz{\'a}lez, Hannam, Husa, Sperhake, and Tichy}}]{BruGonHan06}
\bibinfo{author}{\bibfnamefont{B.}~\bibnamefont{Br{\"u}gmann}},
  \bibinfo{author}{\bibfnamefont{J.~A.} \bibnamefont{Gonz{\'a}lez}},
  \bibinfo{author}{\bibfnamefont{M.}~\bibnamefont{Hannam}},
  \bibinfo{author}{\bibfnamefont{S.}~\bibnamefont{Husa}},
  \bibinfo{author}{\bibfnamefont{U.}~\bibnamefont{Sperhake}}, \bibnamefont{and}
  \bibinfo{author}{\bibfnamefont{W.}~\bibnamefont{Tichy}},
  \bibinfo{journal}{Phys. Rev.} \textbf{\bibinfo{volume}{D77}},
  \bibinfo{pages}{024027} (\bibinfo{year}{2008}), \eprint{gr-qc/0610128}.

\bibitem[{\citenamefont{Husa et~al.}(2008)\citenamefont{Husa, Gonz{\'a}lez,
  Hannam, Br{\"u}gmann, and Sperhake}}]{HusGonHan07}
\bibinfo{author}{\bibfnamefont{S.}~\bibnamefont{Husa}},
  \bibinfo{author}{\bibfnamefont{J.~A.} \bibnamefont{Gonz{\'a}lez}},
  \bibinfo{author}{\bibfnamefont{M.}~\bibnamefont{Hannam}},
  \bibinfo{author}{\bibfnamefont{B.}~\bibnamefont{Br{\"u}gmann}},
  \bibnamefont{and} \bibinfo{author}{\bibfnamefont{U.}~\bibnamefont{Sperhake}},
  \bibinfo{journal}{Class. Quantum Grav.} \textbf{\bibinfo{volume}{25}},
  \bibinfo{pages}{105006} (\bibinfo{year}{2008}), \eprint{arXiv:0706.0740
  [gr-qc]}.

\bibitem[{\citenamefont{Alcubierre and Br{\"u}gmann}(2001)}]{AlcBru00}
\bibinfo{author}{\bibfnamefont{M.}~\bibnamefont{Alcubierre}} \bibnamefont{and}
  \bibinfo{author}{\bibfnamefont{B.}~\bibnamefont{Br{\"u}gmann}},
  \bibinfo{journal}{Phys. Rev. D} \textbf{\bibinfo{volume}{63}},
  \bibinfo{pages}{104006} (\bibinfo{year}{2001}), \eprint{gr-qc/0008067}.

\bibitem[{\citenamefont{Baker et~al.}(2001)\citenamefont{Baker, Br{\"u}gmann,
  Campanelli, Lousto, and Takahashi}}]{BakBruCam01}
\bibinfo{author}{\bibfnamefont{J.}~\bibnamefont{Baker}},
  \bibinfo{author}{\bibfnamefont{B.}~\bibnamefont{Br{\"u}gmann}},
  \bibinfo{author}{\bibfnamefont{M.}~\bibnamefont{Campanelli}},
  \bibinfo{author}{\bibfnamefont{C.~O.} \bibnamefont{Lousto}},
  \bibnamefont{and}
  \bibinfo{author}{\bibfnamefont{R.}~\bibnamefont{Takahashi}},
  \bibinfo{journal}{Phys. Rev. Lett.} \textbf{\bibinfo{volume}{87}},
  \bibinfo{pages}{121103} (\bibinfo{year}{2001}),
  \eprint[http://arXiv.org/abs]{gr-qc/0102037}.

\bibitem[{\citenamefont{Alcubierre et~al.}(2003)\citenamefont{Alcubierre,
  Br{\"u}gmann, Diener, Koppitz, Pollney, Seidel, and Takahashi}}]{AlcBruDie02}
\bibinfo{author}{\bibfnamefont{M.}~\bibnamefont{Alcubierre}},
  \bibinfo{author}{\bibfnamefont{B.}~\bibnamefont{Br{\"u}gmann}},
  \bibinfo{author}{\bibfnamefont{P.}~\bibnamefont{Diener}},
  \bibinfo{author}{\bibfnamefont{M.}~\bibnamefont{Koppitz}},
  \bibinfo{author}{\bibfnamefont{D.}~\bibnamefont{Pollney}},
  \bibinfo{author}{\bibfnamefont{E.}~\bibnamefont{Seidel}}, \bibnamefont{and}
  \bibinfo{author}{\bibfnamefont{R.}~\bibnamefont{Takahashi}},
  \bibinfo{journal}{Phys. Rev. D} \textbf{\bibinfo{volume}{67}},
  \bibinfo{pages}{084023} (\bibinfo{year}{2003}), \eprint{gr-qc/0206072}.

\bibitem[{\citenamefont{Yamamoto et~al.}(2008)\citenamefont{Yamamoto, Shibata,
  and Taniguchi}}]{YamShiTan08}
\bibinfo{author}{\bibfnamefont{T.}~\bibnamefont{Yamamoto}},
  \bibinfo{author}{\bibfnamefont{M.}~\bibnamefont{Shibata}}, \bibnamefont{and}
  \bibinfo{author}{\bibfnamefont{K.}~\bibnamefont{Taniguchi}},
  \bibinfo{journal}{Phys. Rev. D} \textbf{\bibinfo{volume}{78}},
  \bibinfo{pages}{064054} (\bibinfo{year}{2008}).

\bibitem[{\citenamefont{Schnetter et~al.}(2004)\citenamefont{Schnetter, Hawley,
  and Hawke}}]{SchHawHaw03}
\bibinfo{author}{\bibfnamefont{E.}~\bibnamefont{Schnetter}},
  \bibinfo{author}{\bibfnamefont{S.~H.} \bibnamefont{Hawley}},
  \bibnamefont{and} \bibinfo{author}{\bibfnamefont{I.}~\bibnamefont{Hawke}},
  \bibinfo{journal}{Class. Quantum Grav.} \textbf{\bibinfo{volume}{21}},
  \bibinfo{pages}{1465} (\bibinfo{year}{2004}), \eprint{gr-qc/0310042}.

\bibitem[{\citenamefont{Thierfelder et~al.}(2010)\citenamefont{Thierfelder,
  Br\"ugmann, and Galaviz}}]{ThiBruGal10}
\bibinfo{author}{\bibfnamefont{M.}~\bibnamefont{Thierfelder}},
  \bibinfo{author}{\bibfnamefont{B.}~\bibnamefont{Br\"ugmann}},
  \bibnamefont{and} \bibinfo{author}{\bibfnamefont{P.}~\bibnamefont{Galaviz}}
  (\bibinfo{year}{2010}), \eprint{in preparation}.

\bibitem[{\citenamefont{Torigoe et~al.}(2009)\citenamefont{Torigoe, Hattori,
  and Asada}}]{TorHatAsa09}
\bibinfo{author}{\bibfnamefont{Y.}~\bibnamefont{Torigoe}},
  \bibinfo{author}{\bibfnamefont{K.}~\bibnamefont{Hattori}}, \bibnamefont{and}
  \bibinfo{author}{\bibfnamefont{H.}~\bibnamefont{Asada}},
  \bibinfo{journal}{Phys. Rev. Lett.} \textbf{\bibinfo{volume}{102}},
  \bibinfo{pages}{251101} (\bibinfo{year}{2009}).

\bibitem[{\citenamefont{LeVeque}(2007)}]{Lev07}
\bibinfo{author}{\bibfnamefont{R.~J.} \bibnamefont{LeVeque}},
  \emph{\bibinfo{title}{Finite Difference Methods for Ordinary and Partial
  Differential Equations}} (\bibinfo{publisher}{SIAM Press},
  \bibinfo{year}{2007}).

\end{thebibliography}


\end{document}